\documentclass[review,5p,twocolumn]{elsarticle}
\usepackage{lineno,hyperref}
\usepackage{stfloats}
\usepackage{graphicx}
\usepackage{xcolor}
\modulolinenumbers[5]
\usepackage{amsmath,amsfonts}
\usepackage{amssymb}
\usepackage{algorithmic}
\usepackage{algorithm}
\usepackage{array}
\usepackage[caption=false,font=normalsize,labelfont=sf,textfont=sf]{subfig}
\usepackage[font=small,labelfont=bf,labelsep=period,justification=raggedright]{caption}
\usepackage{textcomp}
\usepackage{tikz}
\usepackage{amsthm}
\usepackage{xcolor}
\usepackage{multirow}
\usepackage{nomencl}
\usepackage{tcolorbox}
\usepackage{adjustbox}
\usepackage{booktabs}
\usepackage{pifont}       
\usepackage{bbding}       

\DeclareCaptionLabelFormat{custom}{Fig.\ #2}
\makenomenclature
\usepackage{etoolbox}
\renewcommand{\nomgroup}[1]{%
  \ifstrequal{#1}{A}{}{%
  \ifstrequal{#1}{B}{}{%
  \ifstrequal{#1}{C}{}{}}}%
}

\newtheorem{remark}{Remark}
\newtheorem{proposition}{Proposition}
\biboptions{numbers,sort&compress}
\def\BibTeX{{\rm B\kern-.05em{\sc i\kern-.025em b}\kern-.08em
    T\kern-.1667em\lower.7ex\hbox{E}\kern-.125emX}}
\def\MatrixFont{\bf}
\def\VectorFont{\bf}
\newcommand{\mC}{{\MatrixFont C}}
\newcommand{\mT}{{\MatrixFont \mathbb{T}}}
\newcommand{\mP}{{\MatrixFont P}}
\newcommand{\mF}{{\MatrixFont F}}
\newcommand{\mQ}{{\MatrixFont Q}}
\newcommand{\mR}{{\MatrixFont R}}
\newcommand{\mH}{{\MatrixFont H}}

\newcommand{\mI}{{\MatrixFont I}}
\newcommand{\mM}{{\MatrixFont M}}
\newcommand{\mZ}{{\MatrixFont Z}}
\newcommand{\mS}{{\MatrixFont S}}
\newcommand{\mX}{{\MatrixFont X}}
\newcommand{\mK}{{\MatrixFont K}}

\newcommand{\mE}{{\VectorFont E}}
\newcommand{\vc}{{\VectorFont c}}
\newcommand{\vx}{{\VectorFont x}}

\newcommand{\vz}{{\VectorFont z}}
\newcommand{\vi}{{\VectorFont i}}
\newcommand{\vw}{{\VectorFont w}}
\newcommand{\vf}{{\VectorFont f}}
\newcommand{\vh}{{\VectorFont h}}
\newcommand{\vv}{{\VectorFont v}}

\newcommand{\vs}{{\VectorFont s}}
\definecolor{red}{RGB}{255,0,0}
\definecolor{green}{RGB}{0,125,0}
\definecolor{blue}{RGB}{0,0,255}

\journal{INFORMATION FUSION}

\begin{document}

\begin{frontmatter}

\title{{DIFNet: Decentralized Information Filtering Fusion Neural Network with Unknown Correlation in Sensor Measurement Noises}}


\author[a,b,c]{Ruifeng Dong\fnref{equal}}
\author[d]{Ming Wang\fnref{equal}}
\author[a,b,c]{Ning Liu}
\author[a,b,c]{Tong Guo}
\author[e,f]{Jiayi Kang\corref{cor}}
\ead{kangjiayi@bimsa.cn}
\author[d]{Xiaojing Shen}
\author[a,b,c]{Yao Mao}
\affiliation[a]{organization={State Key Laboratory of Optical Field Manipulation Science and Technology}, 
            city={Chengdu},
            postcode={610209}, 
            state={Sichuan},
            country={P.R.China}}
\affiliation[b]{organization={Institute of Optics and Electronics, Chinese Academy of Sciences},    
            city={Chengdu},
            postcode={610209}, 
            state={Sichuan},
            country={P.R.China}}
\affiliation[c]{organization={University of Chinese Academy of Sciences},     
            city={Beijing},
            postcode={100049}, 
            state={Beijing},
            country={P.R.China}}
\affiliation[d]{organization={School of Mathematics, Sichuan University}, 
            city={Chengdu},
            postcode={610000}, 
            state={Sichuan},
            country={P.R.China}}
\affiliation[e]{organization={Beijing Institute of Mathematical Sciences and Applications}, 
            postcode={101408}, 
            state={Beijing},
            country={P.R.China}}
\affiliation[f]{organization={Hetao Institute of Mathematics and Interdisciplinary Sciences (HIMIS)}, 
            city={Shenzhen},
            postcode={518000}, 
            state={Guangdong},
            country={P.R.China}}
\fntext[equal]{These authors contributed equally to this work}



\cortext[cor]{Corresponding author}


\begin{abstract}
In recent years, decentralized sensor networks have garnered significant attention in the field of state estimation owing to  enhanced robustness, scalability, and fault tolerance. Optimal fusion performance can be achieved under fully connected communication and known noise correlation structures. To mitigate communication overhead, the global state estimation problem is decomposed into local subproblems through structured observation model. This ensures that even when the communication network is not fully connected, each sensor can achieve locally optimal estimates of its observable state components. To address the degradation of fusion accuracy induced by unknown correlations in measurement noise, this paper proposes a data-driven method, termed Decentralized Information Filter Neural Network (DIFNet), to learn unknown noise correlations in data for discrete-time nonlinear state space models with cross-correlated measurement noises. Numerical simulations demonstrate that DIFNet achieves superior fusion performance compared to conventional filtering methods and exhibits robust characteristics in more complex scenarios, such as the presence of time-varying noise.
\end{abstract}


\end{frontmatter}

\linenumbers

\section{Introduction}
Multisensor information fusion refers to the integration of noisy data from multiple sensors or heterogeneous sources to generate more accurate and reliable estimates. Compared to single-sensor systems, multisensor information fusion effectively reduces uncertainty \cite{Trackingand}, enhances system survivability, improves overall reliability and robustness, as well as extends the temporal and spatial coverage of the system \cite{Fortyyearsof}. Recent advancements in wireless sensor networks have driven growing interest in estimation fusion applications across military and civilian fields, e.g., target tracking \cite{bar1993estimation}, aerospace vehicle \cite{Gaussianprocess}, mobile robots \cite{mobilerobots} and autonomous driving \cite{Multimodal}, etc.

Two key issues that need to be considered for the multisensor fusion network to achieve high performance are listed here \cite{Distributeddata}: 
\begin{itemize}
    \item How to choose the appropriate estimation architecture that connects sensors with the processors or agents at the fusion sites and how the data are shared with other sites in the network?
    \item How the processors or agents combine the data from other nodes to provide the best performance? (For example, in centralized systems, how to fuse sensor measurements with dependent measurement errors by a centralized Kalman filter. In distributed systems, the data to be fused may be state estimates with correlated estimation errors due to prior communication, common process noise, correlated measurement noises across sensors, etc. If this dependency is not recognized in fusion, the common information may be double counted, resulting in incorrect fusion results \cite{Areview}.) 
\end{itemize}

\begin{figure*}[!htbp]
\centering
\includegraphics[width = \linewidth]{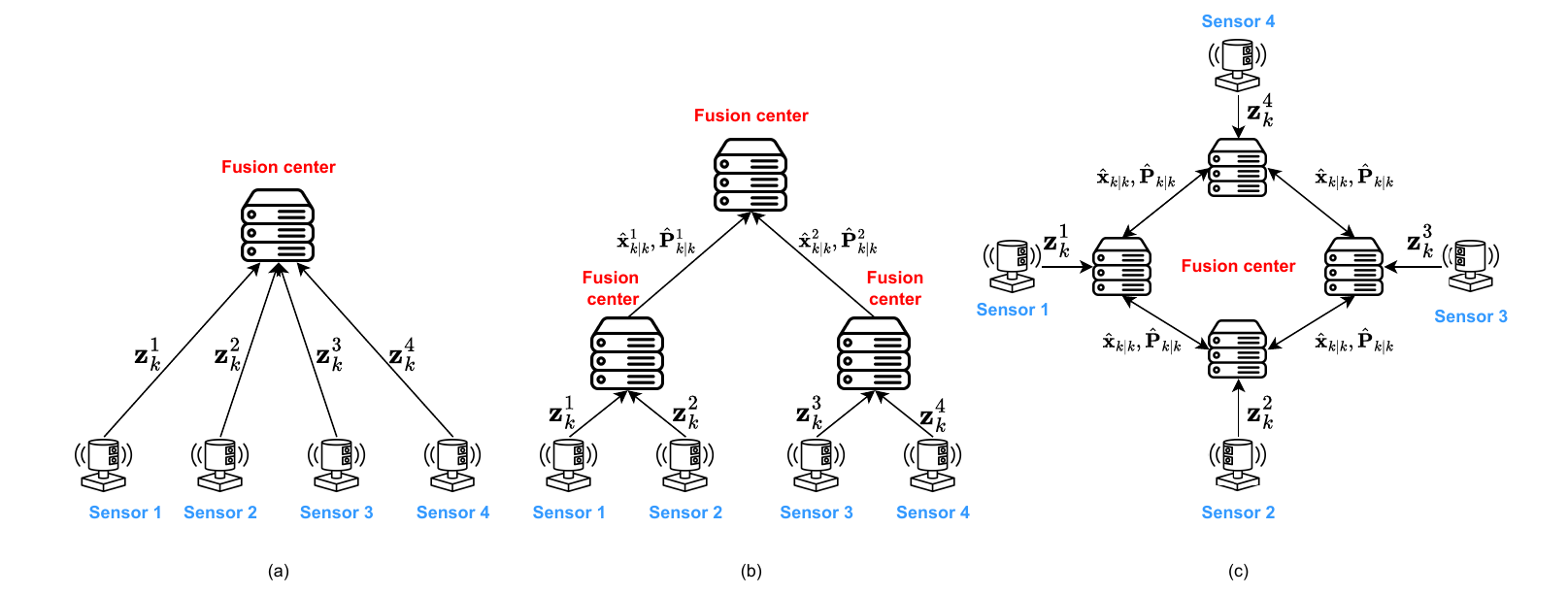}
\label{decentralized}
\caption{Fusion Architecture. The fusion center refers to the processing facilities. (a) Centralized: the fusion center computes the global estimate directly from all sensor measurements $\vz$. (b) Hierarchical: low-level fusion centers compute local estimates $\hat{\vx}, \hat{\mP}$ from their own sensor measurements $\vz$, and a high-level fusion center combines these local estimates into a global estimate. (c) Decentralized: each fusion center computes a local estimate $\hat{\vx}, \hat{\mP}$ from its own measurements $\vz$, and then fuses it with estimates from neighboring centers to produce an individual fusion estimate.}
\label{fusion architecture}
\end{figure*}

For the first issue, three basic fusion architectures are well known \cite{OptimalLinear, Distributeddata}: centralized (corresponding to measurement fusion), hierarchical and decentralized (the latter two are often referred to as distributed, corresponding to estimate fusion), {see Fig. \ref{fusion architecture}.(a).} The centralized architecture comprises a central processor with direct connections to all sensor devices. Each of these devices needs to transmit measurement data to the fusion center and then produce a globally optimal estimate using the expanded measurement. A typical two-level hierarchical architecture is shown in {Fig. \ref{fusion architecture}.(b)}. The low level processors compute state estimates from their local sensor measurements. The high level processor then combines the low level estimates into a global estimate. In fact the most advanced systems today are generally variations of hierarchical structures \cite{Aformally}. The decentralized system comprises a network of sensor devices, each equipped with its own processing facility, thereby eliminating the requirement for centralized processing while maintaining inter-node communication capabilities. Within this system, fusion takes place locally at each node, incorporating local observations and information shared by neighboring nodes. These three fusion architectures have pros and cons in terms of performance, communication, computation, reliability, survivability, etc, (see \cite{Aformally,decentralized}).

\textbf{This paper adopts decentralized fusion architecture for two key motivations.} On the one hand, compared to centralized and hierarchical alternatives, decentralized systems demonstrate superior survivability through fault tolerance with gradual performance reduction when facing failures in communication links, sensors, or processing nodes. They also offer enhanced extensibility characterized by scalability (sensors are easily added or taken away), modularity (much of the code on the processing node is identical, as is the hardware), and flexibility (the number of nodes and how they are connected can easily be varied) \cite{Communications}. On the other hand, to address the theoretical gap in handling unknown noise correlations in decentralized architectures \cite{IFNet}, this paper shifts its focus to decentralized fusion architectures. Together, these two complementary components constitute a comprehensive distributed estimation fusion methodology.

For the second issue, \textbf{this paper focuses on the optimal decentralized fusion with cross-correlated measurement noises (CCMN).} Centralized and hierarchical fusion with CCMN has recently been resolved by model-based scheme, (e.g., cross covariance-based methods \cite{Onthe, Theeffectofthe,MultiSensoroptimal}, covariance intersection method \cite{Anon-divergent,Generaldata,GeneralDecentralized}), hybrid model-based/data-driven scheme \cite{2021KalmanNet,IFNet}. In decentralized systems, various versions of fusion algorithm based on the information form of Kalman filter have been investigated in the literature \cite{Kalmanfilter,Adecentralized,Decentralizedstructures,Fullydecentralised,Towarda,Optimaldecentralized}, with the majority grounded in the assumption of mutual independence among measurement noises. These decentralized estimation algorithms must operate at full communication rates to achieve identical results to their centralized counterparts, which means that communication of information has to be carried out after every measurement \cite{Distributedadaptive, Distributedmultiple}.
If the frequency of communication and fusion is less than the frequency of measurement, the intermittent communication leads to correlated estimation errors due to shared process noise history \cite{Theeffectofthe}, which makes the above-mentioned methods suboptimal. The near-optimal decentralized estimation methods for multisensor information fusion are usually limited by the linear state space assumption and rely on the a priori correlation information of sensor measurement noise at the algorithm implementation level, which poses two key challenges: first, how to construct the optimal fusion strategy under the nonlinear state space model when the correlation of the measurement noise is known; secondly, it is usually difficult to obtain the correlation of the measurement noise, and how to obtain and utilize the correlation information to improve the fusion performance has yet to be explored.

This paper proposes DIFNet, which addresses two fundamental challenges: learning unknown noise correlation structures in nonlinear systems, and maintaining estimation consistency across arbitrary network topologies without requiring full connectivity. The proposed method extends IFNet \cite{IFNet} to decentralized architectures through distributed parameter learning and consensus-based information aggregation. As a hybrid model-based/data-driven approach, it offers stronger interpretability and lower computational demands compared to end-to-end methods. We demonstrate that DIFNet enables consistent estimation in decentralized systems with arbitrary network topologies and can be readily trained via distributed or federated learning schemes. Simulations verify its superior performance over the decentralized information filter (DIF) under process and measurement noise mismatches, both in linear and nonlinear state-space models.

This paper is organized as follows: Section \ref{sec:system_model} presents the model for target tracking problem with CCMN. Then, we introduce model-based decentralized information filter. Section \ref{sec:DIFNet} introduce the model-based data driven method DIFNet. Section \ref{sec:simulations} test the effectiveness of the proposed method in various situations. The main abbreviations and symbols are summarized in the acronyms table and nomenclature table, respectively.

\begin{table}[!htbp]
\centering
\captionsetup{justification=raggedright,singlelinecheck=false,labelformat=empty,margin=0.4cm}
\caption{Acronyms table.}
\scalebox{1.0}{
\begin{tabular}{c|c}
\hline
Acronyms & Full words                           \\ \hline
NN       & Neural network                       \\ \hline
RNN       & Recurrent Neural network                       \\ \hline
FC       & Fully connected                      \\ \hline
GRU      & Gated recurrent unit                 \\ \hline
CCMN     & Cross-correlated measurement noises   \\ \hline
CUMN     & Cross-uncorrelated measurement noises \\ \hline
EKF     & Extended Kalman filter \\ \hline
EIF     & Extended information filter \\ \hline
CI     & Covariance intersection \\ \hline
MSE     & Mean-squared error \\ \hline
RMSE     & Root of the mean-squared error \\ \hline
DIF & Decentralized information filter \\ \hline
DEIF & Decentralized extended information filter \\ \hline
\end{tabular}}
\end{table}

\begin{table}[!htbp]
\centering
\setlength{\tabcolsep}{1.75mm}
\renewcommand{\arraystretch}{1.03}
\scalebox{0.875}{
\begin{tabular}{|ll|}
\hline
\multicolumn{2}{|l|}{Nomenclature table}                                                                                                                                              \\
$\mathbb{R}$                                            & Real number                                                                                                                 \\
$j$                                                     & Sensor index                                                                                                                \\
$(\cdot)^{\mathrm{T}}$                                  & Matrix transpose manipulation                                                                                               \\
$(\cdot)^{\dagger}$                                     & Matrix pseudo-inverse                                                                                                       \\
$k$                                                     & Time instant index                                                                                                          \\
$N$                                                     & Number of sensors                                                                                                           \\
$m$                                                     & Dimension of the system state vector                                          \\
$n_{j}$                                                     &  \begin{tabular}[c]{@{}l@{}}
     Dimension of the measurement vector \\
     of the $j$-th sensor 
\end{tabular}   
\\

$\tilde{n}$ & 
\begin{tabular}[c]{@{}l@{}}
     Dimension of the stacked measurement \\
     vector 
\end{tabular}   

\\
$\mathbf{x}_k \in \mathbb{R}^m$                         & System state vector                                                                                                         \\
$\mathbf{v}_k \in \mathbb{R}^m$                         & Gaussian process noise                                                                                                      \\
$\mathbf{z}_k^j \in \mathbb{R}^{n_{j}}$                       & Measurement vector of the $j$-th sensor                                                            \\
$\mathbf{w}_k^j \in \mathbb{R}^{n_{j}}$                       & Gaussian measurement noise                                                                          \\

$\vf_{k}(\cdot): \mathbb{R}^m \rightarrow \mathbb{R}^m$     & State evolution function                                                                                                    \\
$\mathbf{h}: \mathbb{R}^m \rightarrow \mathbb{R}^{ \tilde{n}}$ & Stacked measurement function                                                                                                \\
$\mathbf{Q}_k \in \mathbb{R}^{m \times m}$              & Covariance of the process noise                                                                                             \\
$\mathbf{R}_k^j \in \mathbb{R}^{n_{j} \times n_{j}}$            & Covariance of $\mathbf{w}_k^j$                                                                                              \\
$\vz_{k} \in \mathbb{R}^{\tilde{n}}$                     & Stacked measurement vector                                                                                                  \\
$\mathbf{w}_k \in \mathbb{R}^{\tilde{n}}$                     & Stacked Gaussian measurement noises                                             \\
$\mathbf{R}_k \in \mathbb{R}^{\tilde{n} \times \tilde{n}}$          & Covariance of $\mathbf{w}_k$                                                                                                \\
$\mathbf{K}_k \in \mathbb{R}^{m \times \tilde{n}}$            & Kalman gain                                                                                                                 \\
$\hat{\mathbf{x}}_{k \mid k-1} \in \mathbb{R}^m$        & Prior estimate of $\mathbf{x}_k$                                                                                            \\
$\mathbf{P}_{k \mid k-1} \in \mathbb{R}^{m \times m}$   & Covariance of $\hat{\mathbf{x}}_{k \mid k-1}$                                                                               \\
$\hat{\mathbf{x}}_{k \mid k} \in \mathbb{R}^m$          & Posterior estimate of $\mathbf{x}_k$                                                                                        \\
$\mathbf{P}_{k \mid k} \in \mathbb{R}^{m \times m}$     & Covariance of $\hat{\mathbf{x}}_{k \mid k}$                                                                                 \\
$\nabla \vf_k \in \mathbb{R}^{m \times m}$       & \begin{tabular}[c]{@{}l@{}}Jacobian matrix of $\vf_k$ evaluated at \\ $\hat{\mathbf{x}}_{k-1 \mid k-1}$\end{tabular} \\
$\nabla \vh_k \in \mathbb{R}^{\tilde{n} \times m}$     & \begin{tabular}[c]{@{}l@{}}Jacobian matrix of $\vh_k$ evaluated at \\ $\hat{\mathbf{x}}_{k \mid k-1}$\end{tabular}   \\
$\triangle \mathbf{z}_k \in \mathbb{R}^{\tilde{n}}$           & \begin{tabular}[c]{@{}l@{}}Difference between predicted measurement \\ and measurement\end{tabular}                       \\
$\mathbf{R}_k^{-1}(* j) \in \mathbb{R}^{\tilde{n} \times n_{j}}$           & The $j$-th column block of $\mathbf{R}_k^{-1}$                                                             \\
$\tilde{\mathbf{z}}_k \in \mathbb{R}^{\tilde{n}}$             & Equivalent linearized measurement vector                                     \\

$\hat{\vi}_{k}^{j} \in \mathbb{R}^{m}$                  & \begin{tabular}[c]{@{}l@{}}
     The $j$-th sensor's local information state \\
     contribution 
\end{tabular}                                                                    \\
$\hat{\mI}_{k}^{j} \in \mathbb{R}^{m \times m}$         & The $j$-th sensor's local information matrix                                                                                \\
$\mM_{k}^{j} \in \mathbb{R}^{m \times m}$               & Fusion weight                                                                                                               \\ \hline
\end{tabular}}
\end{table}

\newpage
\section{SYSTEM MODEL AND PRELIMINARIES} \label{sec:system_model}

\subsection{State Space Model}

We consider a sensor network which contains of $N$ sensors to observe and estimate the state of a moving target. This system is a non-linear, Gaussian and discrete-time state space models which are represented via 
    \begin{subequations}
    \begin{align}
         \vx_{k} &= \vf_{k}(\vx_{k-1}) + \vv_{k} \label{eq:state_transfor},\\
         \vz_{k}^{j} &= \vh_{k}^{j}(\vx_{k}) + \vw_{k}^{j}\label{eq:measurement},
    \end{align}
    \end{subequations}
where $ \vx_{k} \in \mathbb{R}^{m} $ represents the target state vector of the system at time $k$. Then the nonlinear state-evolution function $\vf_{k}(\cdot)$ transfers the state $\vx_{k-1}$ to time $k$. $\vv_{k}$ is white Gaussian noise with a covariance matrix $\mQ_{k}$. Similarly, $\vz_{k}^{j}$ result from the current state vector through a nonlinear measurement mapping $\vh_{k}^{j}(\cdot)$ corrupted by additive white Gaussian noise $\vw_{k}^{j}$ with a covariance matrix $\mR_{k}^{j}$ denoted the vector of measurement at time $k$ from the $j$-th sensor. Many previous researches assumed that $\vw_{k}^{i}$ and $\vw_{k}^{j} $ are unrelated for $i\neq j$, but in this paper, it is noteworthy that $\vw_{k}^{i}$ and $\vw_{k}^{j}$ may exhibit cross-correlation at time instant $k$ which means that cross-covariance matrix $E\left[ \vw_{k}^{i}(\vw_{k}^{j})^{T} \right] = \mR_{k}^{ij} \neq \mathbf{0}$. However in the fusion step of filter, the unknown correlation plays a significant role. To address the challenge of low fusion accuracy arising from this unknown correlation this paper adopts a data-aided scheme.

By stacking the measurement equation, the above multisensor system model can be transformed into the following equation
\begin{subequations}
    \begin{align}
        \vx_{k}&=\vf_{k}(\vx_{k-1})+\vv_{k},\,
\vv_{k}\sim{\mathcal{N}}(\mathbf{0},\mQ_{k}),\,
\,\vx_{k}\in \mathbb{R}^m, \\
\vz_{k}&=\vh_{k}(\vx_{k})+\vw_{k},\,
 \vw_{k}\sim{\mathcal{N}}(\mathbf{0},\mR_{k}),\,
\,\vz_{k}\in \mathbb{R}^{\tilde{n}},
    \end{align}
\end{subequations}
where
\begin{subequations}
\begin{align}
\vz_{k}&=\left[\left(\vz_{k}^{1}\right)^{\rm T},\left(\vz_{k}^{2}\right)^{\rm T},\cdots, \left(\vz_{k}^{N}\right)^{\rm T}\right] ^{\rm T} ,\\
\vh_{k}(\cdot)&=\left[\left(\vh_{k}^{1}(\cdot)\right)^{\rm T},\left(\vh_{k}^{2}(\cdot)\right)^{\rm T},\cdots, \left(\vh_{k}^{N}(\cdot)\right)^{\rm T}\right] ^{\rm T},\\
\vw_{k}&=\left[\left(\vw_{k}^{1}\right)^{\rm T},\left(\vw_{k}^{2}\right)^{\rm T},\cdots, \left(\vw_{k}^{N}\right)^{\rm T}\right] ^{\rm T},
\end{align}
\end{subequations}
\begin{equation}
    \mR_{k}=Cov\left(\vw_{k},\vw_{k} \right) = \left( \begin{matrix}
	\mR_{k}^{11}&		\mR_{k}^{12}&		\cdots&		\mR_{k}^{1N}\\
	\mR_{k}^{21}&		\mR_{k}^{22}&		\cdots&		\mR_{k}^{2N}\\
	\vdots&		\vdots&		\ddots&		\vdots\\
	\mR_{k}^{N1}&		\mR_{k}^{N2}&		\cdots&		\mR_{k}^{NN}\\
\end{matrix} \right) .\label{R} \tag{3d}
\end{equation}
It can be observed that the non-diagonal block of $\mR_{k}$ contain information of cross-correlation $\mR_{k}^{ij}$ which is unknown in practical engineering. For example, unknown cross-correlation may be caused by an electronic jammer.

\subsection{Model-based Multisensor Fusion}
In this subsection, we present a model-based multisensor fusion approach that includes both centralized and decentralized fusion frameworks, where both fusion methods are based on cross-correlated measurement noises.

\subsubsection{Centralized EIF-based Multisensor Fusion}
Centralized EIF-based multisensor fusion is algebraically equivalent to the classical centralized EKF-based multisensor fusion \cite{decentralized}.
Compared with Kalman filtering, information filtering has many advantages, e.g., the centralized information filter has more efficient multisensor fusion with simplified update operations and higher computational efficiency in high-dimensional and sparse systems. Thus, in this paper we choose EIF as our fusion bedrock.

Consistent with Kalman filter, the centralized EIF-based multisensor fusion can be described by a two-step procedure: prediction and update. At each time step $k$, it computes the first and second-order statistical moments.

In the first step, using the system equation in (\ref{eq:state_transfor}) and combining it with the posterior estimate from the previous moment, calculate the prior estimate for this moment as
\begin{subequations}
    \begin{align}
        \hat{\vx}_{k|k-1} &= \vf(\hat{\vx}_{k-1|k-1}),\\
        \mP_{k|k-1} &= \nabla\vf_{k}\cdot\mP_{k-1|k-1}\cdot\nabla\vf_{k}^{T} + \mQ_{k},
    \end{align}\label{predict_measurementfusion}
\end{subequations}
and by utilizing the prior estimation $\hat{\vx}_{k|k-1}$ at this moment, a predicted measurement value $\vh_{k}(\hat{\vx}_{k|k-1})$ can be obtained. Then the innovation $\Delta \vz_{k}$ can be derived as
\begin{equation}
    \Delta \vz_{k} = \vz_{k} - \vh_{k}(\hat{\vx}_{k|k-1}).
\end{equation}
In the update step, the posterior state moments are computed based on the prior moments as
\begin{subequations}
    \begin{align}
        \hat{\vx}_{k|k} &= \mP_{k|k}\left[ \mP_{k|k-1}^{-1}\hat{\vx}_{k|k-1} + \nabla\vh_{k}^{T}\cdot \mR_{k}^{-1}\tilde{\vz}_{k} \right],\\
        \mP_{k|k}^{-1} &= \mP_{k|k-1}^{-1} + \nabla\vh_{k}^{T}\cdot \mR_{k}^{-1}\cdot \nabla \vh_{k},
    \end{align}\label{update_measurementfusion}
\end{subequations}
where $\tilde{\vz}_{k} = \Delta\vz_{k} + \nabla\vh_{k}\cdot\hat{\vx}_{k|k-1}$ is the "equivalent" linearized measurement vector.

From the above equation, it can be seen that when the system is linear Gaussian system, EIF-based centralized fusion is equivalent to IF-based centralized fusion \cite{chong1990distributed} which achieves the optimal performance in the sense of minimizing MSE.

\subsubsection{Model Distribution}

Previous research demonstrates that, under the constraints imposed by the definition of a decentralized sensing network \cite{Datafusion}, it is not in general possible to construct a set of filters which can provide consistent estimates across an arbitrary network topology - with the exception of tree-connected and fully connected networks. The channel filter has been developed to eliminate common information between neighboring nodes \cite{Communicationin,Routingfor,Reliabilityin} (applicable to tree-connected networks), which proves ineffective in general networks with multiple information paths. Fully connected systems, though theoretically viable, suffer from redundant processing and communication that waste computational resources without enhancing survivability. Furthermore, their inherent lack of scalability creates hardware implementation barriers. Model distribution theory \cite{Modeldistribution,decentralized} provides a framework for achieving consistent estimation in decentralized systems with arbitrary network topologies.


Model distribution is the process of constructing reduced order models from a global system model by creating local state vectors which consist of locally relevant states such that there is dynamic equivalence between local and global models. Specifically, the global state $\vx_{k}$ is distributed to $j$-th local state $\vx_{k}^{j}$ via internodal transformation matrix $\mT_{k}^{j}$,
\begin{equation}
    \vx_{k}^{j} = \mT_{k}^{j}\vx_{k}. \label{internodal transformation}
\end{equation}
Note that $\mT_{k}^{j} \in \mathbb{R}^{m_{j}\times m}$ ($m_{j}$ is the dimension of $j$-th sensor's local state space) is defined as any matrix that arbitrarily picks states or combinations of states from the global state $\vx_{k}$. In the following contents, we assume that the rank of $\mT_{k}^{j}$ is no less than $n_{j}$, i.e., the dimension of $j$-th sensor measurement (follows that $n_{j} \leq rank(\mT_{k}^{j}) \le m$), which produces model size reduction without redundant states in the local state vector.

Based on \eqref{internodal transformation}, the local state space model can be represented as
\begin{subequations}
    \begin{align}
        \vx_{k}^{j}&= \vf_{k}^{j}(x_{k-1}^{j}) + \vv_{k}^{j},\\
        \vz_{k}^{j}&= \vc_{k}^{j}(\vx_{k}^{j}) + \vw_{k}^{j}.
    \end{align}
\end{subequations}
Here $\vv_{k}^{j}=\mT_{k}^{j}\vv_{k}$ is the process noise vectors related to the global noise vector, $\vf_{k}^{j}(\cdot)$ and $\vc_{k}^{j}(\cdot)$ is the local state evolution function and local observation function, respectively, which satisfy
\begin{subequations}
\begin{align}
        \vf_{k}^{j}(\mT_{k-1}^{j}\vx_{k-1})&=\mT_{k}^{j}\vf_{k}(\vx_{k-1}),\\
    \vh_{k}^{j}(\vx_{k})&=\vc_{k}^{j}(\mT_{k}^{j}\vx_{k}).
\end{align}
\end{subequations}
In the linear system, the notations could be simplified consistent with \cite{Modeldistribution}, e.g.,
\begin{subequations}
    \begin{align}
        \vx_{k}^{j}&= \mF_{k}^{j}\vx_{k} + \vv_{k}^{j}, \label{nonlinear nodal transition}\\
        \vz_{k}^{j}&= \mC_{k}^{j}\vx_{k}^{j} + \vw_{k}^{j}.
    \end{align}
\end{subequations}
where
\begin{subequations}
    \begin{align}
        \mF_{k}^{j}&=\mT_{k}^{j} \mF_{k}\mT_{k-1}^{j\dagger},\label{local evolution function}\\
        \mC_{k}^{j}&=\mH_{k}^{j}\mT_{k}^{j \dagger}.
    \end{align}
\end{subequations}
Here, the superscript $``\dagger"$ stands for Moore-Penrose pseudo-inverse. \eqref{local evolution function} can be obtained by pre-multiplying the global state transition equation (the evolution function $\vf_{k}(\cdot)$ reduces to the transition matrix $\mF_{k}$ in the linear system) by $\mT_{k}^{j}$ and substituting into the nodal state transition equation \eqref{nonlinear nodal transition} the time appropriate forms of the state space relation \eqref{internodal transformation}. 

\subsubsection{Decentralized EIF-based Multisensor Fusion with Cross-Correlated Measurement Noises}
In cases where the sensor measurement noises $\vw_{k}^{j}$ are cross-correlated, decentralized EIF-based multisensor fusion with CCMN can be obtained by the local EKF step, communication step, fusion step and feedback step.

\textbf{Local EKF:} At the time step $k$, all local sensors perform their individual EKF, for example, for the $j$-th sensor, it firstly performs the following prediction step,
\begin{subequations}
\label{local EKF predict}
\begin{align}
\hat{\vx}_{k|k-1}^{j}&=\vf_{k}^{j}\left( \hat{\vx}_{k-1|k-1}^{j} \right), \\ 
\mP_{k|k-1}^{j}&=\nabla \vf_{k}^{j}\cdot\mP_{k-1|k-1}^{j}\cdot\nabla (\vf_{k}^{j})^{\rm T}+\mQ_{k}^{j}, \\
\hat{\vz}_{k|k-1}^{j}&=\vc_{k}^{j}\left( \hat{\vx}_{k|k-1}^{j} \right) ,\\
\mS_{k|k-1}^{j}&=\nabla \vc_{k}^{j} \cdot \mP_{k|k-1}^{j}\cdot(\nabla \vc_{k}^{j})^{\rm T}+\mR_{k}^{j},\label{S}
\end{align}
\end{subequations}
where the $j$-th sensor's local Jacobian matrix is
\begin{subequations}
    \begin{align}
        \nabla \vf_{k}^{j}&=\mT_{k}^{j}\cdot\nabla \vf_{k}\cdot\mT_{k-1}^{j\dagger},\\
        \nabla \vc_{k}^{j}&=\nabla \vh_{k}^{j}\cdot\mT_{k}^{j\dagger}. \label{c}
    \end{align}
\end{subequations}
Then the local estimates and error covariances are calculated by
\begin{subequations}
\label{local EKF update}
\begin{align}
\hat{\vx}_{k|k}^{j}&=\hat{\vx}_{k|k-1}^{j}+\mK_{k}^{j}(\vz_{k}^{j}-\hat{\vz}_{k|k-1}^{j}) ,\,\,\label{updatex}\\
\mP_{k|k}^{j}&=\mP_{k|k-1}^{j}-\mK_{k}^{j}\mS_{k|k-1}^{j}(\mK_{k}^{j})^{\rm T}.\label{P1}
\end{align}
\label{update}
\end{subequations}
\hspace{-7pt} Here, $\mK_{k}^{j}$ is the $j$-th sensor's Kalman gain, it is given by
\begin{equation}
\mK_k^{j}=\mP_{k|k-1}^{j}\cdot(\nabla\vc_{k}^{j})^{\rm T}\cdot(\mS_{k|k-1}^{j})^{-1}.\label{K1}
\end{equation}
\textbf{Communication \footnote{In hierarchical architecture, we reduce communication load by sending $\hat{\vx}_{k|k}^{j}$ and $\mP_{k|k-1}^{j}$ to the fusion center, which, however, needs to store these local information.}:} All sensors send their local estimates and error covariances $\hat{\vx}_{k|k-1}^{j}$, $\hat{\vx}_{k|k}^{j}$, $\mP_{k|k-1}^{j}$, $\mP_{k|k}^{j},\, j=1,2,\cdots, N$  to their neighbours (interested nodes whose state subspace $\vx_{k}^{i}$ are related).

\textbf{Fusion:} At the fusion step, upon receiving local communications, the nodes combine these estimates through the update step. In the update step (\ref{global DIF with CCMN}-\ref{local fusion weight}), the posterior moments are computed based on the local information state contribution  $\hat{\vi}_{k}^{j}$ and its associated local information matrix $\hat{\mI}_{k}^{j}$ \cite{Hierarchicalestimation}.

\textbf{Feedback:} The fusion center feeds back the fused estimate $\hat{\vx}_{k|k}^{j}$ and error covariance $\mP_{k|k}^{j}$ to each local EKF.

\begin{proposition}
\label{propostion1} 
When measurement noises $\vw_{k}^{j}$ are cross-correlated, $\nabla \vh_{k}^{j}$ is of full row rank ($rank(\nabla \vh_{k}^{j})=n_{j}$) and the rank of $\mT_{k}^{j}$ is no less than $n_{j}$, the decentralized EIF-based multisensor fusion with CCMN given by the following (\ref{global DIF with CCMN}-\ref{global fusion weight}) is optimal in the sense of being equivalent to the centralized EIF-based multisensor fusion (\ref{predict_measurementfusion}-\ref{update_measurementfusion}).  Meanwhile, the local estimates given by (\ref{local DIF with CCMN}-\ref{local fusion weight}) are consistent for each node in the sensor network.
\begin{itemize}
    \item Global space: 
    \begin{subequations}
    \label{global DIF with CCMN}
    \begin{align}
    &\hat{\vx}_{k \mid k} =\mP_{k \mid k}[\mP_{k \mid k-1}^{-1}\hat{\vx}_{k \mid k-1}+\sum_{j=1}^N \mM_{k}^{j}(\mT_{k}^{j})^{\rm T}\hat{\vi}_k^j] ,\label{global DIF with CCMN_x}\\
    &\mP_{k \mid k}^{-1}=\mP_{k \mid k-1}^{-1}+\sum_{j=1}^N \mM_{k}^{j}(\mT_{k}^{j})^{\rm T}\hat{\mI}_k^j\mT_{k}^{j},\label{global DIF with CCMN_P}
    \end{align}
    \end{subequations}
    where
    \begin{equation}
        \mM_{k}^{j}=\nabla \vh_{k}^{\rm T}\cdot\mR_{k}^{-1}\left( \ast j \right) \mR_{k}^{j}\cdot\left( \nabla \vh_{k}^{j^\dagger} \right) ^{\rm T}. \label{global fusion weight}
    \end{equation}
     Here, $\mR_{k}^{-1}\left( \ast j \right)$ denotes the $j$-th column block of $\mR_{k}^{-1}$.
    \item Local subspace:
    \begin{subequations}
    \label{local DIF with CCMN}
    \begin{align}
    &\hat{\vx}_{k \mid k}^{i} =\mP_{k \mid k}^{i}[(\mP_{k \mid k-1}^{i})^{-1}\hat{\vx}_{k \mid k-1}^{i}+\sum_{j=1}^N \tilde{\mM}_{k}^{j}(\mT_{k}^{ij})^{\rm T}\hat{\vi}_k^j] ,\label{local DIF with CCMN_x}\\
    &(\mP_{k \mid k}^{i})^{-1}=(\mP_{k \mid k-1}^{i})^{-1}+\sum_{j=1}^N \tilde{\mM}_{k}^{j}(\mT_{k}^{ij})^{\rm T}\hat{\mI}_k^j\mT_{k}^{ij},\label{local DIF with CCMN_P}
    \end{align}
    \end{subequations}
    where fusion weights switch to 
    \begin{equation}
        \tilde{\mM}_{k}^{j}=(\mT_{k}^{i\dagger})^{\rm T}\cdot \nabla \vh_{k}^{\rm T}\cdot\mR_{k}^{-1}\left( \ast j \right) \mR_{k}^{j}\cdot\left( \nabla \vh_{k}^{j^\dagger} \right) ^{\rm T}\cdot(\mT_{k}^{i})^{\rm T}. \label{local fusion weight}
    \end{equation}
\end{itemize}
where
\begin{subequations}
\label{iI}
\begin{align}
&\hat{\vi}_{k}^{j}=\left(\mP_{k|k}^{j}\right)^{-1}\hat{\vx}_{k|k}^{j}-\left(\mP_{k|k-1}^{j}\right)^{-1}\hat{\vx}_{k|k-1}^{j},\label{i} \\
&\hat{\mI}_{k}^{j}=\left(\mP_{k|k}^{j}\right)^{-1}-\left(\mP_{k|k-1}^{j}\right)^{-1}.  \label{I}
\end{align}
\end{subequations}
The internodal transformation matrix $\mT_{k}^{ij}$ operating the informaiton from sensors $i$ to $j$, summarized by
\begin{equation}
     \mT_{k}^{ij} = \mT_{k}^{j}(\mT_{k}^{i \dagger}).\label{Tij}
\end{equation}
Proof: see Appendix A.
\end{proposition}
Proposition \ref{propostion1} is established under when the row dimensions of all sensor measurement matrices $\nabla \vh_{k}^{j}$ are less than or equal to the dimension of the state, and all $\nabla \vh_{k}^{j}$ to be of full row rank, i.e., $\nabla \vh_{k}^{j}(\nabla \vh_{k}^{j})^{\dagger}=\mE_{n_{j}\times n_{j}}$ ($\mE_{n_{j}\times n_{j}}$ is the identity matrix), such assumption is fulfilled very often in the tracking problem \cite{zhu2012networked}. 
\begin{remark}
    When dynamic system is the linear system, i.e., evolution function $\vf_{k}(\cdot)$ reduce to matrix $\mF_{k}$ and observation function $\vh_{k}^{j}(\cdot)$ reduce to matrix $\mH_{k}^{j}$, the decentralized IF-based multisensor fusion with CCMN shares an identical form with the fusion formula  (\ref{global DIF with CCMN}) and (\ref{iI}) in the nonlinear system. 
\end{remark}
\begin{remark}
Communication: This depends on connectedness, which in turn depends on internodal transformations. Two nodes $i$ and $j$ will communicate if, and only if, they have an overlapping information space, that is, at least one of the internodal transformation matrices, $\mT_{k}^{ij}$ and $\mT_{k}^{ji}$ is not a null matrix. When communication does take place, only relevant information is exchanged.
\end{remark}
The internodal transformation matrix $\mT_{k}^{ij}$ determines the configuration of the nodal network. It indicates which nodes are connected and what
they communicate as well as giving the relationships between nodal models.
Consider the following connection and communication scenarios. 
\begin{itemize}
    \item  If sensor $i$ has no information relevant to $j$, then $\mT_{k}^{ij}=(\mT_{k}^{ji})^{\rm T} = \mathbf{0}_{n_{j}n_{i}}$, the Null matrix. 
    \item When $\mT_{k}^{ij} = \mathbf{I}$ for all $i, j = 1,2,\cdots,N$ then all nodes have the same (global) model and the problem reduces to decentralized fully connected network.
    \item When the network topology is distributed but centralized (hierarchical architecture), i.e nodes with different models communicate with a central processor but not with each other, then there is only one $\mT_{k}^{ij} = \mT_{k}^{i} = \mT_{k}^{iG}$ at each node $i$. 
    \item Arbitrary tree, ring or loop connected topologies based on internodal transformations are possible. In fact, the fully connected decentralized, the hierarchical and the centralized control configurations are special cases of the distributed decentralized control network where for all nodes: {$\mT_{k}^{ij} = \mI$}, {$\mT_{k}^{ij} = \mT_{k}^{iG}$} and {$\mT_{k}^{ij} = \mathbf{0}$}, respectively.
\end{itemize}

\begin{remark}
    When the rank of $\mT_{k}^{j}$ is less than $n_{j}$, i.e., the dimension of $j$-th sensor measurement, (this is usual in practical systems, e.g., decoupling coupled multisensor states or reducing model order) the decentralized EIF-based multisensor fusion with CCMN fails to achieve estimation consistency equivalent to centralized fusion. Nevertheless, this trade-off intentionally sacrifices partial fusion accuracy to significantly alleviate inter-sensor communication loads, which constitutes the foundational objective of model distribution theory.
\end{remark}

The equations of the decentralized EIF-based multisensor fusion with CUMN or CCMN are summarized in \ref{tab:equations}. While the original derivation of decentralized multisensor fusion with CUMN in \cite{Modeldistribution} addressed linear systems, here we extend it to nonlinear systems while preserving its structural equivalence. It's worth noting that fusion with CCMN is identical to fusion with CUMN except for additional weights that depend on the cross-correlation.

\begin{table*}[!t]
\renewcommand{\arraystretch}{1.5}
\setlength{\tabcolsep}{3mm}{}
\caption{\vspace{-2pt} Equations of the decentralized EIF-based multisensor fusion with CUMN or CCMN.}
\scalebox{0.86}{
\begin{tabular}{cc|cl|c}
\hline
\multicolumn{4}{c|}{Global Space}                                    & Local Subspace\\ \hline
\multicolumn{1}{c|}{\multirow{3}{*}{CUMN}} & Prediction: & \multicolumn{2}{c|}{$\begin{array}{c}
\hat{\vx}_{k|k-1}=\vf_{k}\left( \hat{\vx}_{k-1|k-1} \right),\\
\mP_{k|k-1}=\nabla \vf_{k}\cdot\mP_{k-1|k-1}\cdot\nabla \vf_{k}^{\rm T}+\mQ_{k}.
\end{array}$} & {$\begin{array}{c}
\hat{\vx}_{k|k-1}^{i}=\vf_{k}^{i}\left( \hat{\vx}_{k-1|k-1}^{i} \right),\\
\mP_{k|k-1}^{i}=\nabla \vf_{k}^{i}\cdot\mP_{k-1|k-1}^{i}\cdot\nabla (\vf_{k}^{i})^{\rm T}+\mQ_{k}^{i}.
\end{array}$}                \\ \cline{2-5} 
\multicolumn{1}{c|}{}                      & Update:     & \multicolumn{2}{c|}{$\begin{array}{c}
\hat{\vx}_{k \mid k} =\mP_{k \mid k}[\mP_{k \mid k-1}^{-1}\hat{\vx}_{k \mid k-1}+\sum_{j=1}^N (\mT_{k}^{j})^{\rm T}\hat{\vi}_k^j] ,\\
\mP_{k \mid k}^{-1}=\mP_{k \mid k-1}^{-1}+\sum_{j=1}^N (\mT_{k}^{j})^{\rm T} \hat{\mI}_{k}^{j}\mT_{k}^{j}.
\end{array}$} & {
$\begin{array}{c}
\hat{\vx}_{k \mid k}^{i} =\mP_{k \mid k}^{i}[(\mP_{k \mid k-1}^{i})^{-1}\hat{\vx}_{k \mid k-1}^{i}+\sum_{j=1}^N (\mT_{k}^{ij})^{\rm T}\hat{\vi}_k^j] ,\\
(\mP_{k \mid k}^{i})^{-1}=(\mP_{k \mid k-1}^{i})^{-1}+\sum_{j=1}^N (\mT_{k}^{ij})^{\rm T} \hat{\mI}_{k}^{j}\mT_{k}^{ij}.
\end{array}$}                \\ \hline
\multicolumn{1}{c|}{\multirow{3}{*}{CCMN}} & Prediction: & \multicolumn{2}{c|}{$\begin{array}{c}
\hat{\vx}_{k|k-1}=\vf_{k}\left( \hat{\vx}_{k-1|k-1} \right),\\
\mP_{k|k-1}=\nabla \vf_{k}\cdot\mP_{k-1|k-1}\cdot\nabla \vf_{k}^{\rm T}+\mQ_{k}.
\end{array}$} & {$\begin{array}{c}
\hat{\vx}_{k|k-1}^{i}=\vf_{k}^{i}\left( \hat{\vx}_{k-1|k-1}^{i} \right),\\
\mP_{k|k-1}^{i}=\nabla \vf_{k}^{i}\cdot\mP_{k-1|k-1}^{i}\cdot\nabla (\vf_{k}^{i})^{\rm T}+\mQ_{k}^{i}.
\end{array}$}                \\ \cline{2-5} 

\multicolumn{1}{c|}{}                      & Update:     & \multicolumn{2}{c|}{$\begin{array}{c}
\hat{\vx}_{k \mid k} =\mP_{k \mid k}[\mP_{k \mid k-1}^{-1}\hat{\vx}_{k \mid k-1}+\sum_{j=1}^N\mM_{k}^{j}(\mT_{k}^{j})^{\rm T}\hat{\vi}_{k}^{j}],\\
\mP_{k \mid k}^{-1}=\mP_{k \mid k-1}^{-1}+\sum_{j=1}^N\mM_{k}^{j}(\mT_{k}^{j})^{\rm T}\hat{\mI}_{k}^{j}\mT_{k}^{j}.
\end{array}$} & {$\begin{array}{c}
\hat{\vx}_{k \mid k}^{i} =\mP_{k \mid k}^{i}[(\mP_{k \mid k-1}^{i})^{-1}\hat{\vx}_{k \mid k-1}^{i}+\sum_{j=1}^N \tilde{\mM}_{k}^{j}(\mT_{k}^{ij})^{\rm T}\hat{\vi}_k^j] ,\\
(\mP_{k \mid k}^{i})^{-1}=(\mP_{k \mid k-1}^{i})^{-1}+\sum_{j=1}^N \tilde{\mM}_{k}^{j}(\mT_{k}^{ij})^{\rm T} \hat{\mI}_{k}^{j}\mT_{k}^{ij}.
\end{array}$}                \\ \hline
\end{tabular}}
\label{tab:equations}
\end{table*}

\begin{figure*}[t!]
\centering
\includegraphics[width=\textwidth]{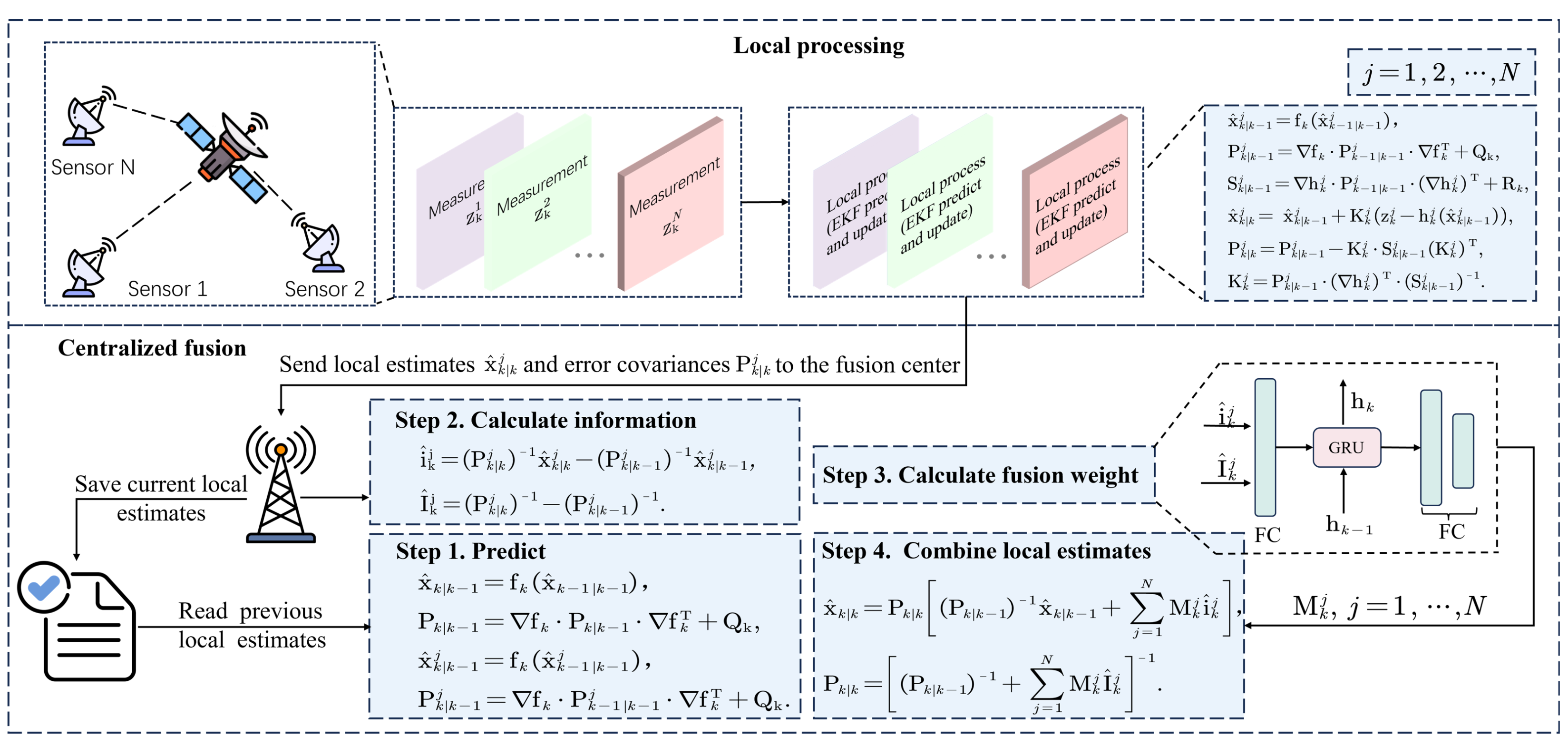}
\caption{Schematic diagram of the proposed DIFNet. The top part shows the generation of local estimates and error covariances by local sensors. The bottom part illustrates the communication and implementation flow at each fusion center.}
\label{schematic diagram}
\end{figure*}

\section{DIFNet}\label{sec:DIFNet}
\subsection{Motivation and Distributed Architecture Design}

The theoretical framework presented in Section 2 demonstrates optimal performance under ideal conditions. However, practical applications encounter significant challenges. Traditional model-based fusion requires precise knowledge of the cross-correlation structure $\mR^{ij}_{k}$, which is often unknown or time-varying. Computing the fusion weights $\mM^{j}_{k}$ involves complex matrix operations that become computationally intensive in large-scale networks, with communication overhead potentially scaling quadratically with the number of sensors. To overcome these limitations, we propose DIFNet, inspired by federated learning paradigms.

This distributed architecture provides several key advantages: enhanced communication efficiency through local processing and effcient parameter sharing, fault tolerance against single points of failure, privacy preservation by retaining sensitive data at individual nodes, and inherent scalability for dynamic sensor networks. Unlike centralized approaches that rely on sharing raw measurements, our framework enables sensors to collaboratively learn optimal fusion strategies while preserving data locality. Unlike existing distributed fusion methods that require explicit cross-correlation modeling, our approach learns these relationships implicitly through neural networks. This federated learning inspiration is reflected in three core principles:
\begin{enumerate}
    \item[(1)] Local model training, where each sensor maintains a neural network to learn local fusion parameters that contribute to global fusion strategies.
    \item[(2)] Parameter aggregation, achieved by exchanging learned network parameters rather than raw measurements.
    \item[(3)] Iterative improvement, allowing online adaptation to changing environmental conditions.
\end{enumerate}

\subsection{Network Architecture and Training Strategy}

Firstly, we use the same input-output configuration which has been validated for its effectiveness in \cite{IFNet} as IFNet of DIFNet. To learn the parameter $\mM_{k}^{j}$ which related to $\mR_{k}^{ij}$, we take $\hat{\vi}_{k}^{j}$ and $\hat{\mI}_{k}^{j}$ as inputs to DIFNet, where
\begin{equation*}
    \begin{aligned}
        \hat{\vi}_{k}^{j} &= (\mP_{k|k}^{j})^{-1}\hat{\vx}_{k|k}^{j} - (\mP_{k|k-1}^{j})^{-1}\hat{\vx}_{k|k-1}^{j},\\
        \hat{\mI}_{k}^{j} &= (\mP_{k|k}^{j})^{-1} - (\mP_{k|k-1}^{j})^{-1}.
    \end{aligned}
\end{equation*}

\begin{figure}[htbp]
\centering
\includegraphics[width=\linewidth]{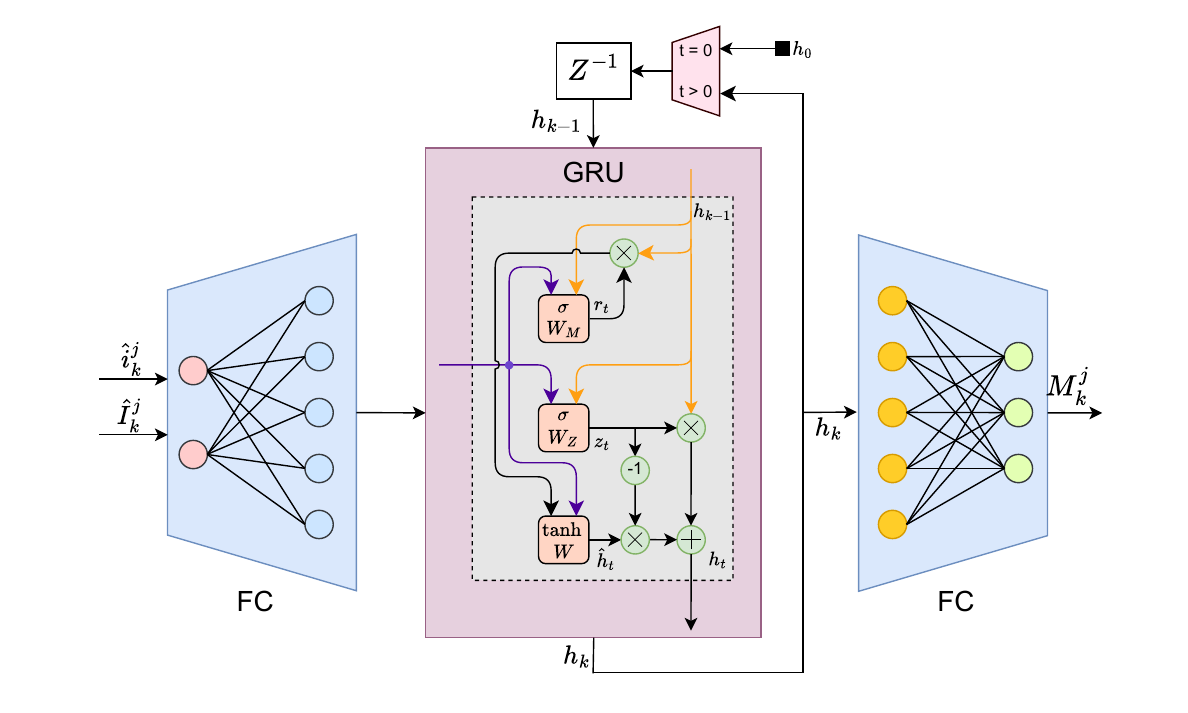}
\caption{RNN block of DIFNet in each sensor, consisting of FC and GRU.}
\label{fig:RNN_block}
\end{figure}

In order to better solve the problem of time-series data, we chose RNN as the backbone of NN architecture instead of a FC Network. Moreover, RNN has also been successfully applied in filtering algorithms for continuous filtering systems \cite{YauYau2025}. To further address the gradient vanishing and exploding of RNN, here we used GRU owing to its low complexity and popularity. Firstly, we set up a FC layer followed by ReLU activation maps input features to high-dimensional space. The input of the FC layer is sized as the dimension of $\hat{\vi}_{k}^{j}$ and $\hat{\mI}_{k}^{j}$ which is $(m+m^2)N$, where $m$ is the dimension of the system state vector and $N$ is the number of sensors. If there is no communication between sensors, all corresponding inputs will be set to 0. The output of the FC layer serves as the input to GRU. The hidden state of the GRU is sized as an integer product of $m^2$. Then the GRU state vector is mapped to another space , utilizing an FC layer whose output dimension is also an integer product of $m^2$. Finally, we use a output FC layer whose output dimension is $m^2N$ to estimate $\mM_{k}^{j}, \enspace j=1,\cdots,N$.  The network physical architecture diagram that implements GRU is illustrated in Fig.\ref{fig:RNN_block}.

As mentioned before, each sensor maintains a neural network to learn fusion weights, so there are a total of $N$ neural networks in the sensor network. At each time instance $k$, $j$-th sensor first performs an EKF locally, and then determines whether it needs to communicate with $i$-th sensor based on $\mT_{k}^{ij}$, for $i \neq j$. If communication is required, $j$-th sensor transmits $\hat{\vx}_{k|k-1}^{j}, \mP_{k|k-1}^{j}, \hat{\vx}_{k|k}^{j}$ and $\mP_{k|k}^{j}$ are to the $i$-th sensor and receives information from the $i$-th sensor. Afterwards, $j$-th sensor calculate the input of its neural network $\hat{\vi}_{k}^{i}$ and $\hat{\mI}_{k}^{i}$ locally, and the fusion weights $\mM_{k}^{ij}$ are estimated using the neural network. Finally, based on the obtained fusion weights, the received information is fused with the local state estimation to obtain the final state estimation.

For the training algorithm, let a dataset $\mathcal{D} = \left\{ (\mX^{l},\mZ^{1,l},\cdots,\mZ^{N,l}) \right\}_{l=1}^{L}$ consists of $L$ different target trajectories and corresponding measurement data from each sensors. Let $T_{l}$ be the length of $l$-th trajectory and $\Theta$ be the trainable parameter of NN. The training of the neural networks utilizes labeled $\mathcal{D}$. For the NN of the $j$-th sensor, we calculate the loss function $\mathcal{L}_{j}$ based on the ground truth state $\vx_{k}$ which differs from the estimated state $\hat{\vx}_{k|k}^{j}$ of the $j$-th sensor. For networks training, we use MSE as the loss function,
\begin{equation}
    \mathcal{L}_{j}(\Theta) = \frac{1}{L}\sum\limits_{l=1}^{L}\frac{1}{T_{l}}\sum\limits_{k=1}^{T_{l}}||\vx_{k}^{l} - \hat{x}_{k|k}^{l,j}\left(\Theta\right)||_{2}^{2} + \gamma||\Theta||_{2}^{2},
\end{equation}
where $\hat{\vx}_{k|k}^{l,j}$ denotes the estimated state of the $j$-th sensor for the $l$-th trajectory at time k, and $\gamma > 0$ is a regularization coefficient used to avoid network overfitting. To
optimize the parameter $\Theta$, we adopt the Adam optimizer, combined with a CyclicLR scheduler to adjust the learning rate.

Unlike traditional methods that depend on explicit correlation matrices, DIFNet addresses cross-correlated measurement noise (CCMN) implicitly through two mechanisms:
\begin{enumerate}
    \item[(1)] Correlation-aware feature extraction to identify patterns indicating cross-correlation.
    \item[(2)] Temporal modeling that incorporates historical information to capture time-varying correlations.
\end{enumerate}
This data-driven approach offers inherent robustness to unknown correlation patterns, parameter drift, and sensor degradation, eliminating the need for predetermined models.


\section{Experiments} \label{sec:simulations}
In this section we will conduct numerical simulation experiments with cross-correlated measurement noises to verify the effectiveness and feasibility of DIFNet.

In out experimental study, we will compare DIFNet with Decentralized EIF-based estimate fusion with CCMN which can be considered as representing the ideal performance, Decentralized EIF-based estimate fusion with CUMN, and Centralized EIF-based estimate fusion with CCMN. The estimation accuracy of different sensors is evaluated by Root Mean Square Error (RMSE) calculated in the local state space of each sensor. RMSE of $j$-th sensor at time instant $k$ is defined as follows:
\begin{equation*}
    \mathrm{RMSE}_{k}^{j} = \sqrt{\frac{1}{N}\sum\limits_{i=1}^{N}|| \mT_{k}^{j}\vx_{k,i} - \mT_{k}^{j}\hat{\vx}_{k,i}^{j} ||_{2}^{2}},
\end{equation*}
where $\vx_{k,i}$ and $\hat{\vx}_{k,i}^{j}$ are the target state and estimation state at time k of the $i$-th experiment, respectively.

\subsection{Experimental setting}
In our experiments, DIFNet is tested in the sensor network containing multiple sensor nodes with varying measurement information and measurement noise, all measurements of sensors are cross-correlated.

The dataset used for training contains 100 trajectories, the dataset used for cross validation contains 20 trajectories, and the dataset used for testing contains 40 trajectories, each of which contains 50 time steps. The optimizer for neural network training is Adam, with a fixed learning rate of $10^{-3}$, batch size of 20, and weight decay of $10^{-4}$.

\subsection{Linear state space model}\label{sec:linear_state_space}
In the example, we consider a constant velocity motion system where the state of the target is $\left[ x,\dot{x},y,\dot{y},z,\dot{z} \right]^{T}$. Teh state transition matrix and process noise covariance matrix of the target are set as
\begin{equation*}
    \mF_{k} = \left( \begin{matrix}
        1 & T & 0 & 0 & 0 & 0\\
        0 & 1 & 0 & 0 & 0 & 0\\
        0 & 0 & 1 & T & 0 & 0\\
        0 & 0 & 0 & 1 & 0 & 0\\
        0 & 0 & 0 & 0 & 1 & T\\
        0 & 0 & 0 & 0 & 0 & 1
    \end{matrix} \right),
\end{equation*}
\begin{equation*}
    \mQ_{k} = q^2 * \mI_{3\times3} \otimes \left( \begin{matrix}
        \frac{T^4}{3} & \frac{T^3}{2} \\
        \frac{T^3}{2} & T^2
    \end{matrix}\right),
\end{equation*}
where the sampling interval T = 1s and the process noise standard deviation q = 1. The $\otimes$ is Kronecker product.

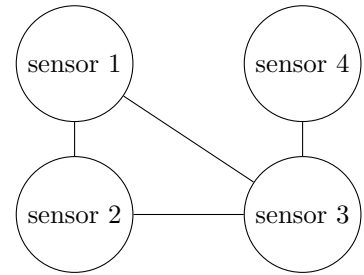
\begin{figure}[htbp]
\centering
\begin{tikzpicture}[scale=1]
    \node[circle, draw, minimum size=1cm] (s1) at (0, 2) {sensor 1};
    \node[circle, draw, minimum size=1cm] (s2) at (0, 0) {sensor 2};
    \node[circle, draw, minimum size=1cm] (s3) at (3, 0) {sensor 3};
    \node[circle, draw, minimum size=1cm] (s4) at (3, 2) {sensor 4};
    \draw (s1) -- (s2);
    \draw (s1) -- (s3);
    \draw (s2) -- (s3);
    \draw (s4) -- (s3);
\end{tikzpicture}
\caption{ The comminication relationship of 4 sensors in sensor network.}
\label{fig:linear_sen4_communication}
\end{figure}

We consider the sensro network consisting of 4 different sensors whose measurement valuse are $\left[ x,\dot{x},y,\dot{y} \right]^T, \left[ x,\dot{x},y,\dot{y} \right], \left[ x,\dot{x},y,\dot{y},z,\dot{z} \right]$ and $\left[ z,\dot{z} \right]$, respectively. Therefore the internodal transformation matrix are $\mT_{k}^{1} = \left[ \mI_{4\times4}, \mathbf{0}_{4\times2} \right], \mT_{k}^{2} = \left[ \mI_{4\times4},\mathbf{0}_{4\times2} \right], \mT_{k}^{3} = \mI_{6\times6}$ and $\mT_{k}^{4} = \left[ \mathbf{0}_{2\times4}, \mI_{2\times2} \right]$. Thus the communication relationship between sensors is shown in the Fig.\ref{fig:linear_sen4_communication}.

We assume that the natural measurement noise of the 4 sensors $\vw_{k}^{j}$, for $j$=1,2,3,4, are mutually independent. However there is a jammer signal $\vw_{k}^{0}$ with covariance $\mR_{k}^{0}$ which will affect each sensor in sensor network. Thus the actual measurement noise at $j$-th sensor is given by
\begin{equation*}
    \tilde{\vw}_{k}^{j} = \vw_{k}^{j} + \beta_{j}\mT_{k}^{j}\vw_{k}^{0},
\end{equation*}
where the $\beta_{j}$ is a constant used to reflect the degree of influence of jammer noise on $j$-th sensor. Thus the stacked noise covariance $\tilde{\mR}_{k}$ can be computed by
\begin{equation*}
    \tilde{\mR}_{k}^{ij} = \begin{cases}
        \beta_{i}\beta_{j} \mT_{k}^{i} \mR_{k}^{0} (\mT_{k}^{j})^T, &  \rm{if} \quad i \neq j \\
        \mR_{k}^{i} + \beta_{i}\beta_{j} \mT_{k}^{i} \mR_{k}^{0} (\mT_{k}^{j})^T, & \rm{if}  \quad i=j.
    \end{cases}
\end{equation*}
where $\mR_{k}^{i}$ is the covariance of $\vw_{k}^{j}$. In this case we set $\beta_{j}=0.5$, for $j=1,2,3,4$. We set
$$\mR_{k}^{0} = \rm{diag}\left[ 100^2,10^2,100^2,10^2,100^2,10^2 \right],$$
$$\mR_{k}^{1} = \rm{diag}\left[ 100^2,10^2,100^2,10^2\right],$$
$$\mR_{k}^{2} = \rm{diag}\left[ 200^2,20^2,200^2,20^2 \right],$$
$$\mR_{k}^{3} = \rm{diag}\left[ 200^2,20^2,200^2,20^2,200^2,20^2\right],$$
$$\mR_{k}^{4} = \rm{diag}\left[ 100^2,10^2,100^2,10^2 \right].$$

The initial state of target $\vx_{0}$ is set as $\left[ 0,100,0,100,0,100 \right]^T$. The initial state estimation $\hat{\vx}_{0|0}$ is set as $\left[ 
100,100,100,100,100,100 \right]^{T}$ and $\mP_{0|0} = 10000 * \mI_{6\times6}$. In the case Decentralized IF-inexact, DIFNet employ partially inaccurate filter parameters, i.e., $q=5, \tilde{\mR}_{k} = \rm{diag}\left[ 
\sqrt{\mR_{k}^{1}},\sqrt{\mR_{k}^{1}},\sqrt{\mR_{k}^2},\sqrt{\mR_{k}^{3}} \right]$ where $\sqrt{\cdot}$ represents the square root of each element in the matrix. Centralized KF and Decentralized IF-exact employ accurate filter parameters.

\begin{figure}[htbp]
    \centering
    \includegraphics[width = \linewidth]{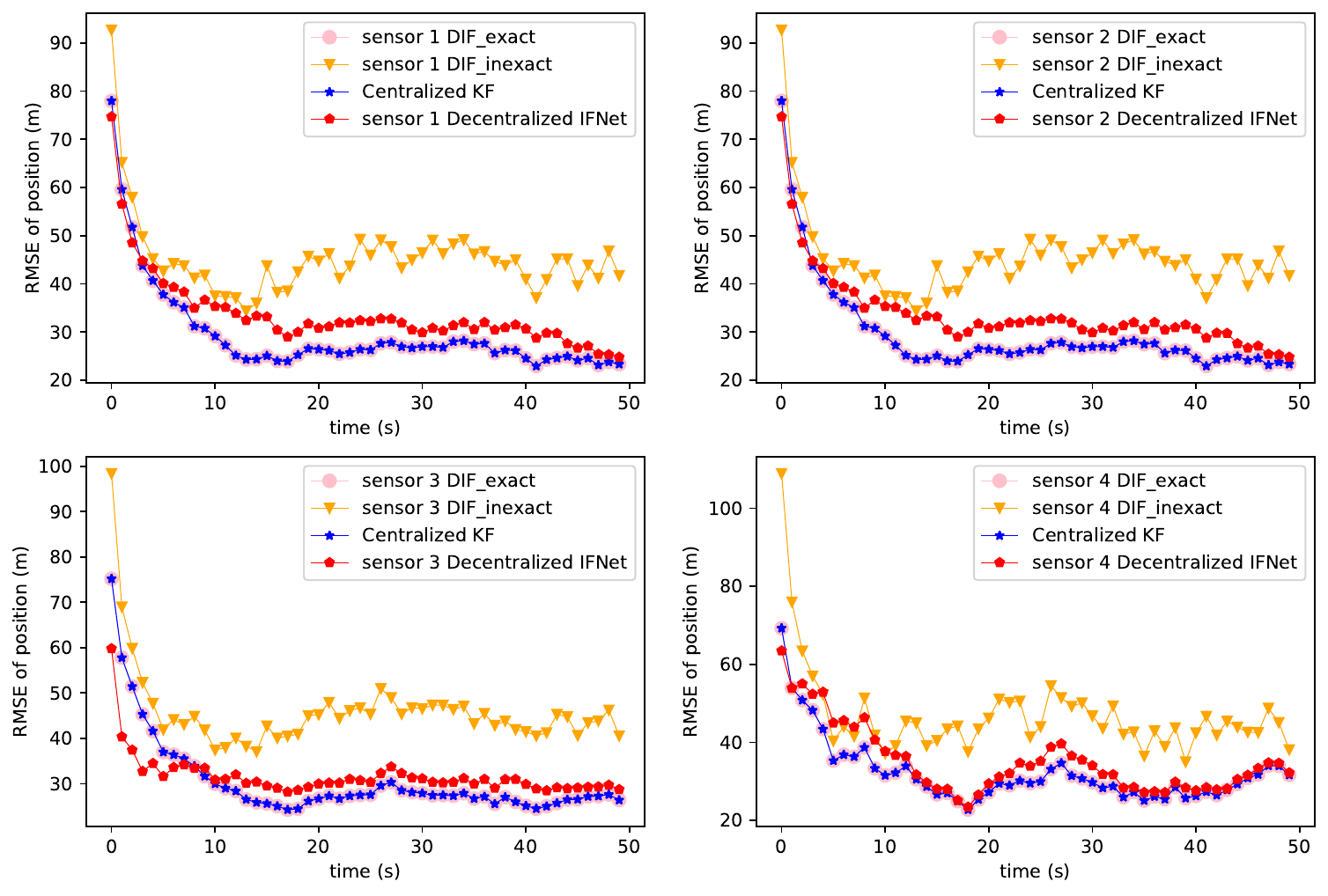}
    \caption{RMSE of estimated position under linear state space model, tested on the test dataset.}
    \label{fig:linear_s4_rmse_pos}
\end{figure}

\begin{figure}[htbp]
    \centering
    \includegraphics[width=\linewidth]{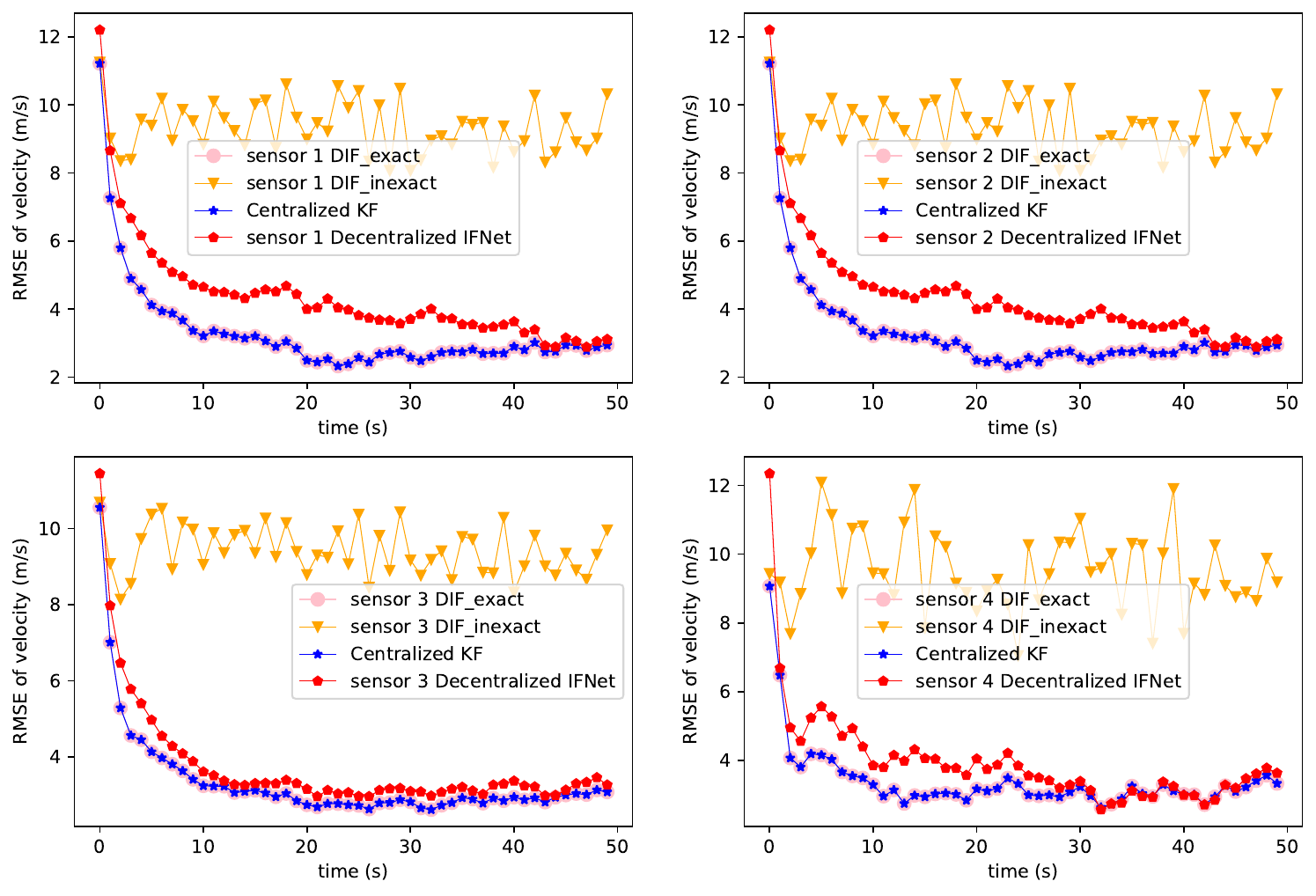}
    \caption{RMSE of estimated velocity under linear state space model, tested on the test dataset}
    \label{fig:linear_s4_rmse_val}
\end{figure}

The RMSE of the different methods is presented in Fig.\ref{fig:linear_s4_rmse_pos} and Fig.\ref{fig:linear_s4_rmse_val}. First, we can see that the performance of Centralized KF and Decentralized IF-exact is completely consistent, which is in line with expectations. This indicates that compared to fully connected communication networks, DIF reduces communication burden without sacrificing accuracy. The results also indicate that DIFNet outperforms all filtering algorithms except for those with precise models. This also directly indicates that the neural network has successfully learned the fusion weight matrix $\mM_{k}^{j}$ from the input data.

In addition, since the initialization parameters of the neural networks deployed on sensor 1 and sensor 2 are consistent during the training process, and the training inputs and errors are also the same, the final training results of the two neural networks are also the same.

\subsection{Nolinear state space model}\label{sec:nonliear_state_space_model}
In this experiment, we consider a target with state $\left[ x,\dot{x},y,\dot{y},z,\dot{z} \right]^{T}$ which has a constant speed and $\dot{x}$ and $\dot{y}$ have a constant known turn rate. That is to say, the state transition matrix and process noise covariance matrix of the target are set as (\ref{eq:nonlinear_F}) and (\ref{eq:nonlinear_Q}). In this example we set $\omega = 0.05$, $T=1$ and $q=1$. 

We consider a sensor network which contains 4 sensors. The first sensor measures the azimuth angle and the horizontal velocity of the target. The second sensor measures the horizontal distance between the target and itself, as well as the horizontal velocity of the target. The third sensor directly measures  states of the target. The last sensor directly measures the last two dimensions of the target state. According to different information measured by different sensors, their respective $\mT_{k}$ are $\mT_{k}^{1} = \left[ 
\mI_{4\times4}, \mathbf{0}_{4\times2} \right]$, $\mT_{k}^{2} = \left[ \mI_{4\times4}, \mathbf{0}_{4\times2} \right] $, $\mT_{k}^{3} = \mI_{6\times6} $ and $\mT_{k}^{4} = \left[ \mathbf{0}_{2\times4}, \mI_{2\times2} \right]$. Thus the communication relationship between the 4 sensors is consistent with that shown in Fig.\ref{fig:linear_sen4_communication}.
\begin{figure*}[htbp]
\begin{equation}\label{eq:nonlinear_F}
    \mF_{k} = \left(
    \begin{matrix}
        1 & \frac{\sin\omega T}{\omega} & 0 & -\frac{1 - \cos\omega T}{\omega} 
 & 0 & 0\\
        0 & \cos\omega T & 0 & -\sin\omega T &0 & 0 \\
        0 & \frac{1 - \cos\omega T}{\omega} & 1 & \frac{\sin\omega T}{\omega} & 0 & 0 \\
        0 & \sin\omega T & 0 & \cos\omega T & 0 & 0\\
        0 & 0 & 0 & 0 & 1 & T\\
        0 & 0 & 0 & 0 & 0 & 1
\end{matrix}
    \right),
\end{equation}
\begin{equation}\label{eq:nonlinear_Q}
    \mQ = q^2 \left(\begin{matrix}
\frac{2(\omega T - \sin\omega T)}{\omega^{3}} & \frac{1 - \cos\omega T}{\omega^{2}} & 0 & \frac{\omega T - \sin\omega T}{\omega^{2}} & 0 & 0 \\
\frac{1 - \cos\omega T}{\omega^{2}} & T & -\frac{\omega T - \sin\omega T}{\omega^{2}} & 0 & 0 & 0\\
0 & -\frac{\omega T - \sin\omega T}{\omega^{2}} & \frac{2(\omega T - \sin\omega T)}{\omega^{3}} & \frac{1 - \cos\omega T}{\omega^{2}} & 0 & 0 \\
\frac{\omega T - \sin\omega T}{\omega^{2}} & 0 & \frac{1 - \cos\omega T}{\omega^{2}} & T & 0 & 0\\
 0 & 0 & 0 & 0 & \frac{T^4}{3} & \frac{T^3}{2} \\
 0 & 0 & 0 & 0 & \frac{T^3}{2} & T^2
        \end{matrix}\right)
\end{equation}
\end{figure*}

In this example, we also consider that besides natural measurement noise of each sensor $w_{k}^{j}$, for $j=1,2,3,4$, there exists a jammer signal $\vw_{k}^{0}$ with covariance $\mR_{k}^{0}$. We set $\mR_{k}^{0}$ as 
\begin{equation*}
\begin{aligned}
    \mR_{k}^{0}
    = \mathrm{diag}\left[ (1^{\circ})^2, 150^2,15^2,100^2,10^2,100^2,10^2,100^2,10^2   \right].
\end{aligned}
\end{equation*}
Therefore, the actual measurement noise at $j$-th sensor is given by
\begin{equation*}
    \tilde{\vw}_{k}^{j} = \vw_{k}^{j} + \beta_{j}\tilde{\mT}_{k}^{j}\vw_{k}^{0},
\end{equation*}
where
\begin{equation*}
    \begin{aligned}
        \tilde{\mT}_{k}^{1} &= \left( \begin{matrix}
            1 & 0 & 0 & 0 & 0 & 0 & 0 & 0 & 0 \\
            0 & 0 & 1 & 0 & 0 & 0 & 0 & 0 & 0
        \end{matrix} \right), \\
        \tilde{\mT}_{k}^{2} &= \left(\begin{matrix}
            0 & 1 & 0 & 0 & 0 & 0 & 0 & 0 & 0 \\
            0 & 0 & 1 & 0 & 0 & 0 & 0 & 0 & 0
        \end{matrix}\right), \\
        \tilde{\mT}_{k}^{3} &= \left[ \mathbf{0}_{6\times3}, \mI_{6\times6} \right], \\
        \tilde{\mT}_{k}^{4} &= \left[ \mathbf{0}_{2\times7}, \mI_{2\times2} \right].
    \end{aligned}
\end{equation*}
The parameter $\beta_{j}$, for $j=1,2,3,4$, represents the degree of
influence of jammer noise on $j$-th sensor. The calculation of the stacked noise covariance $\tilde{\mR}_{k}$ here is the same as the linear state space model, except that $\mT_{k}^{j}$ is replaced with $\tilde{\mT}_{k}^{j}$.  We set $\beta_{j} = 2$ and 
\begin{equation*}
    \begin{aligned}
        \mR_{k}^{1} &= \mathrm{diag}\left[ (1^{\circ})^2,15^2 \right],\\
        \mR_{k}^{2} &= \mathrm{diag}\left[ 250^2,25^2 \right], \\
        \mR_{k}^{3} &= \mathrm{diag}\left[ 200^2,20^2,200^2,20^2,200^2,20^2 \right], \\
        \mR_{k}^{4} &= \mathrm{diag}\left[ 100^2,10^2 \right].
    \end{aligned}
\end{equation*}

The initial state of target $\vx_{0}$ is set as $\left[ 0,100,0,100,0,100 \right]^T$. The position of 4 sensors are set as
\begin{equation*}
    \begin{aligned}
        \vs_{0} &= \left[ -5500,1000,0 \right]^T,\\
        \vs_{1} &= \left[ -5000,0,0 \right]^T,\\
        \vs_{2} &= \left[ 500,300,0 \right]^T,\\
        \vs_{4} &= \left[ 50,500,0 \right]^T.
    \end{aligned}
\end{equation*}
The initial state estimation $\hat{\vx}_{0|0}$ is set as $\left[ 275,10,275,10,275,10 \right]^T$ and $\mP_{0|0} = 100^2 * \mI_{6\times6}$. Decentralized EIF-inexact and DIFNet employ partially inaccurate filter parameters, i.e., $q=5,\tilde{R}_{k} = \rm{diag}\left[ \sqrt{\mR_{k}^{1}},\sqrt{\mR_{k}^{2}},\sqrt{\mR_{k}^{3}},\sqrt{\mR_{k}^{4}} \right]$. Centralized EKF and Decentralized EIF-exact employ accurate filter parameters.

\begin{figure}[htbp]
    \centering
    \includegraphics[width = \linewidth]{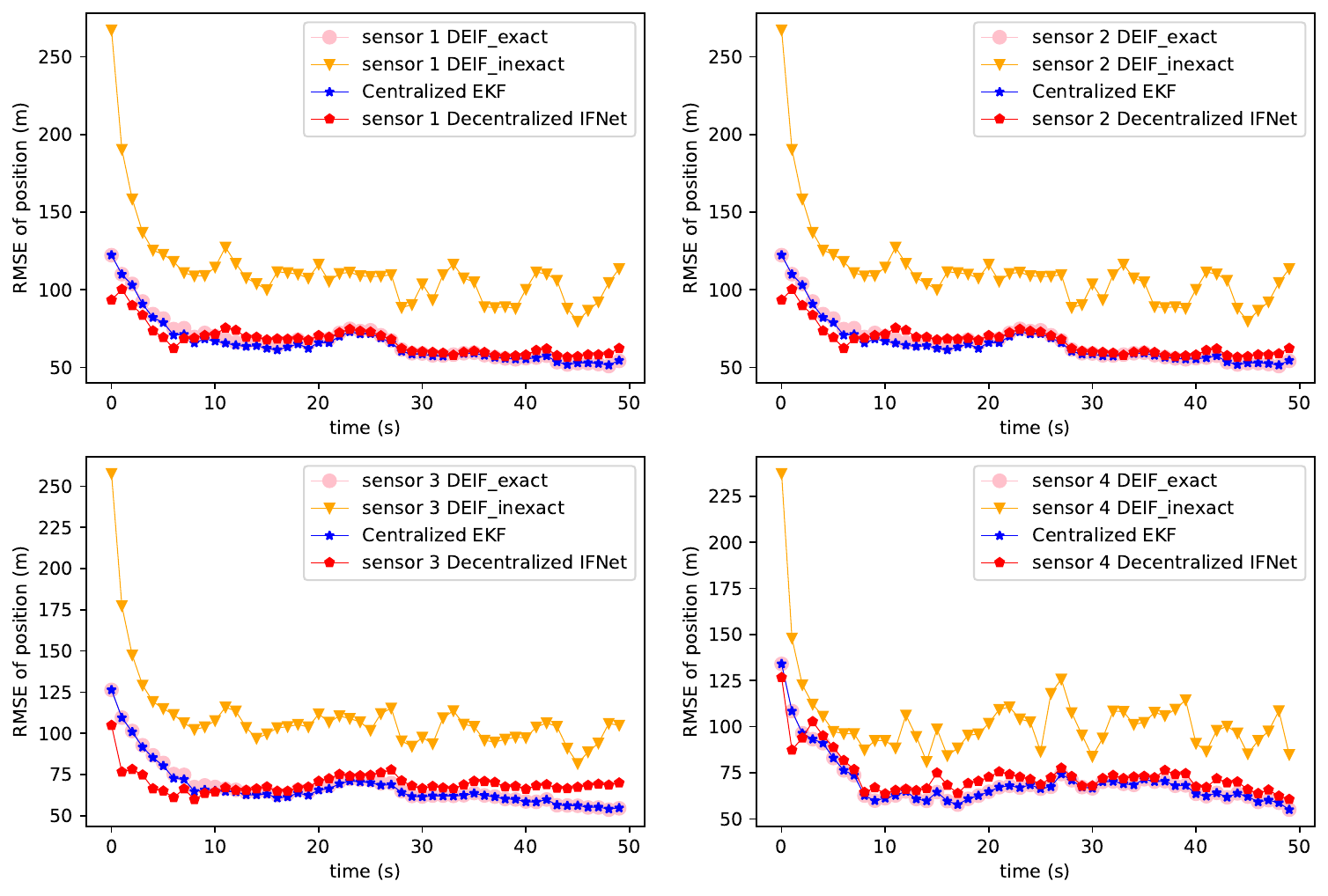}
    \caption{RMSE of estimated position under nolinear state space model, tested on the test dataset.}
    \label{fig:Nolinear_s4_rmse_pos}
\end{figure}

\begin{figure}[htbp]
    \centering
    \includegraphics[width=\linewidth]{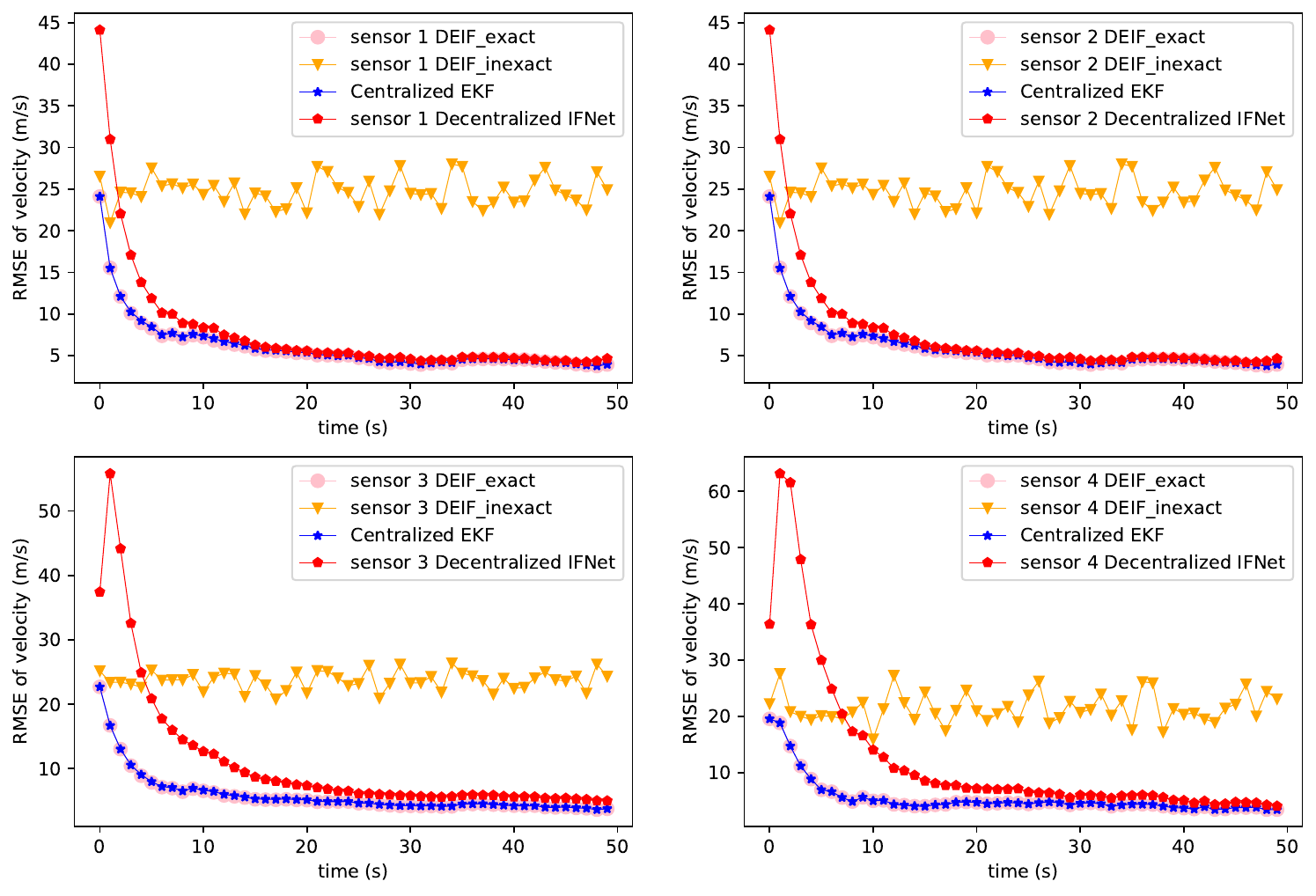}
    \caption{RMSE of estimated velocity under nolinear state space model, tested on the test dataset}
    \label{fig:Nolinear_s4_rmse_val}
\end{figure}

The results of the simulation experiment are shown in Fig.\ref{fig:Nolinear_s4_rmse_pos} and Fig.\ref{fig:Nolinear_s4_rmse_val}. We can find that DIFNet still can achieve a better results than the DIF method with imprecise parameters and have a similar performance with DIF method with precise parameters in the case of nonlinear measurement functions. 

\subsection{Time-varying noise}
\begin{figure}[htbp]
    \centering
    \includegraphics[width=\linewidth]{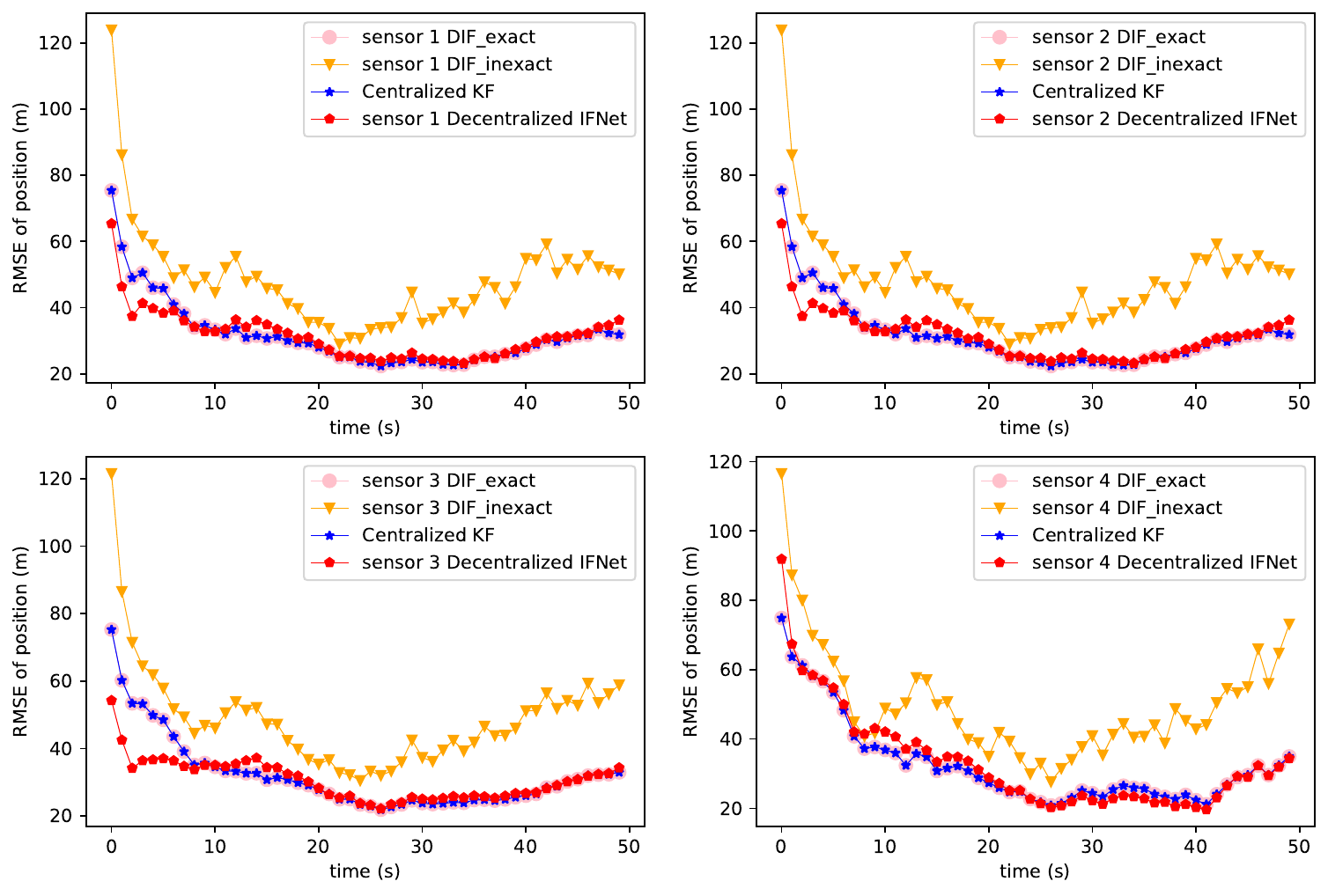}
    \caption{RMSE of estimated position under time-varying ($\sigma=0.5$) measurement noise case, tested on the test dataset.}
    \label{fig:time_var_linear_s4_pos_rmse}
\end{figure}

\begin{figure}[htbp]
    \centering
    \includegraphics[width=\linewidth]{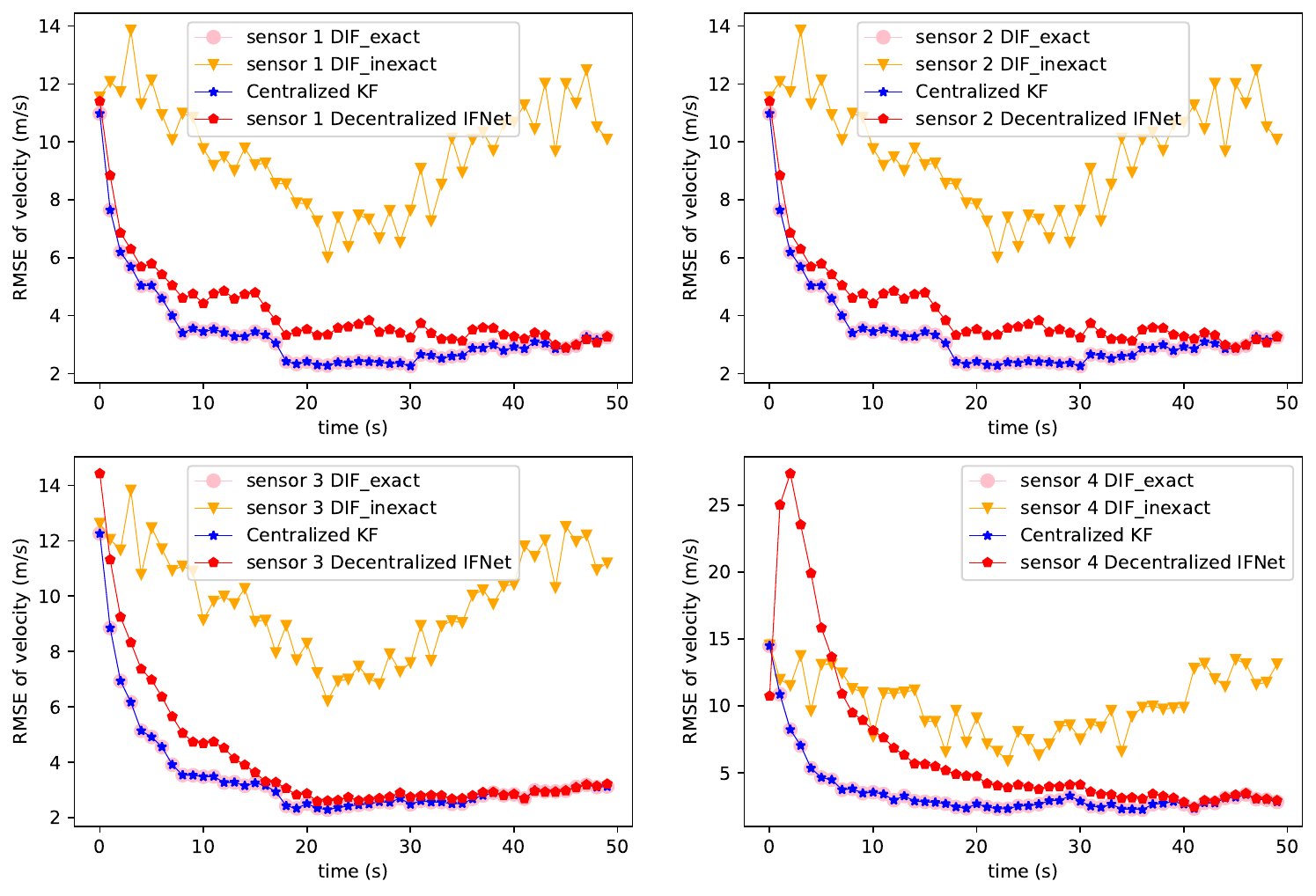}
    \caption{RMSE of estimated velocity under time-varying ($\sigma=0.5$) measurement noise case, tested on the test dataset.}
    \label{fig:time_var_linear_s4_vel_rmse}
\end{figure}

In practice, noise not only has unknown correlations, but it may also be time-varying. In this part, we will test DIFNet under time-varying measurement noise in linear state space model presented in \ref{sec:linear_state_space}. We assume that the covariance matrix $\mR_{k}^{var}$ under time-varying case at time instance $k$ satisfies the following equation
\begin{equation*}
    \mR_{k}^{var} = (1+\sigma\cos(2\pi\frac{k}{T})\tilde{R}_{k},
\end{equation*}
where $\tilde{R}_{k}$ which is unknown to the sensor is the  measurement noise covariance matrix in linear state space model, $\sigma \in (-1,1)$ is a parameter that reflects the time-varying degrees of measurement noise. Here, we let $\sigma = 0.5$, and the remaining experimental settings are the same as linear state space model. The results are displayed in Fig.\ref{fig:time_var_linear_s4_pos_rmse}
 and Fig.\ref{fig:time_var_linear_s4_vel_rmse}.

From Fig.\ref{fig:time_var_linear_s4_pos_rmse} and Fig.\ref{fig:time_var_linear_s4_vel_rmse}, we can observe that DIFNet can still achieve similar results to DIF with precise model parameters (except $\sigma$), and performs better in estimating velocity than in the linear state space model. This indicates that the new method successfully learns the fusion weight matrix $\mM_{k}^{j}$ even in the presence of time varying measurement noise.

\begin{figure}[htbp]
    \centering
    \includegraphics[width=\linewidth]{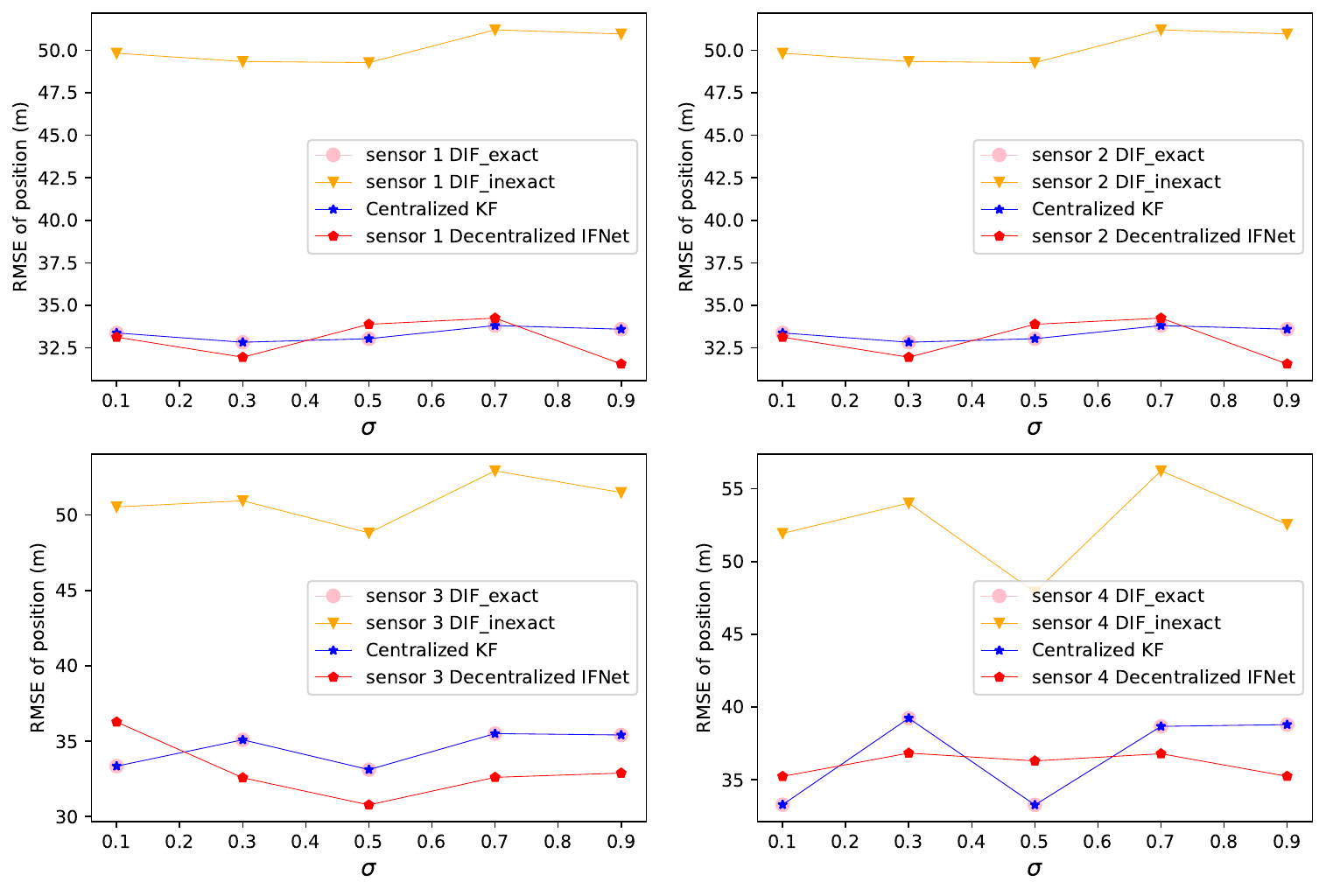}
    \caption{RMSE of estimated position under different values of $\sigma$, tested on the test dataset.}
    \label{fig:var_rmse_pos}
\end{figure}

\begin{figure}[htbp]
    \centering
    \includegraphics[width=\linewidth]{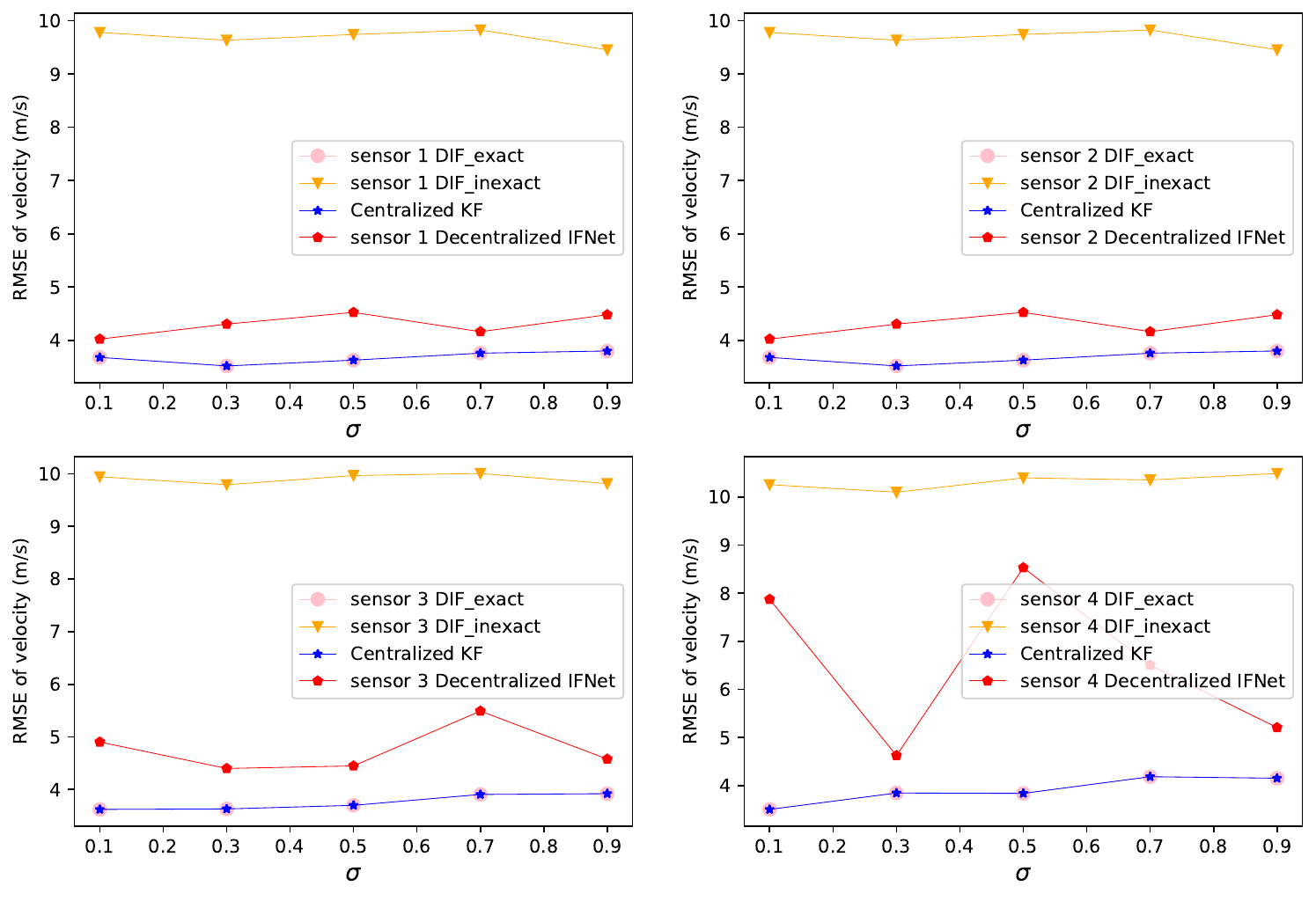}
    \caption{RMSE of estimated velocity under different values of $\sigma$, tested on the test dataset.}
    \label{fig:var_rmse_vel}
\end{figure}

In order to further investigate the robustness of DIFNet under different time-varying noise, we test it under with different values of $\sigma$, calculated the RMSE of different methods, and plotted them. The results are displayed in Fig.\ref{fig:var_rmse_pos} and Fig.\ref{fig:var_rmse_vel}.

From Fig.\ref{fig:var_rmse_pos} and Fig.\ref{fig:var_rmse_vel}, we can observe that under different values of $\sigma$, DIFNet can always successfully learn the fusion weight matrix $\mM_{k}^{j}$. Due to the unknown parameter $\sigma$ for each sensor, the model of each filtering method is not completely accurate. In this case, DIFNet can sometimes perform better even than DIF where all parameters except $\sigma$ are accurately known. DIFNet is designed for cross-correlated measurement noises not time-varying noises, but this numerical experiment also shows that even when encountering unknown time-varying measurement noise, DIFNet can perform much better with the same imprecise model parameters as DIF\_inexact. This also indicates thar DIFNet has sufficient robustness to ensure that it can maintain good performance even in unexpected situations encountered in practical applications.

\subsection{Time consumptions analysis}
In this section, we demonstrate the average time consumptions for different methods to track a single trajectory under the linear and nonlinear state space model presented in \ref{sec:linear_state_space} and \ref{sec:nonliear_state_space_model}, respectively. Since the only difference between DIF and DIFNet lies in data fusion step, in order to obtain a more accurate comparison, we only compare the time consumption of different methods in the data fusion step. All the numerical experiments are performed on a laptop with Intel(R) Core(TM) i7-14650HX CPU @
2.20GHZ. To enable intuitive comparison of computational efficiency, the execution time of the DIF\_exact  is taken as the reference (normalized to 1 unit). The reported values represent the ratio of each method’s elapsed time to that of the traditional method. The experimental results are shown in Table.\ref{tab:time_consimptions}. The reported values shows that the time cost of DIFNet is acceptable.

\begin{table}[htbp]
\caption{The time consumptions of data fusion steps for different filter methods to track a single trajectory, take the execution time of DIF\_exact as the reference, tested on the test dataset.}
\begin{tabular}{cccc}
\hline
State space model & DIF\_exact  & DIFNet \\ \hline
Linear            & 1              & 2.5256            \\
Nonlinear         & 1               & 2.7628              \\ \hline
\end{tabular}
\label{tab:time_consimptions}
\end{table}

\section{Conclusion}\label{sec:conclusion}
This paper proposes a novel neural network-based approach that learns optimal fusion weights for the DIF framework without requiring explicit knowledge of noise cross-correlations. The key innovation lies in learning the mapping from local information matrices to optimal fusion weights $\mM_{k}^{j}$ circumventing the need for explicit covariance matrix computation. The proposed DIFNet architecture enables distributed parameter learning across sensor networks while maintaining theoretical convergence guarantees under mild connectivity conditions.  Simulation results illustrate the  the effectiveness of DIFNet takes on both linear and nonlinear systems with CCMN. The proposed method demonstrates superior performance compared to conventional DIF approaches, particularly under model uncertainties such as time-varying noise characteristics and parameter mismatches. The framework opens several avenues for future research, including unsupervised online adaptation mechanisms and extension  to heterogeneous sensor networks with asynchronous measurements. Theoretical analysis of convergence properties and bounds under different network conditions represents another promising direction for investigation.

\newpage
\appendix
\section{Proof of the decentralized EIF-based fusion with cross-correlated measurement noises}
The proof starts from the global state of decentralized EIF-based fusion with CCMN \eqref{global DIF with CCMN} derived from the centralized EIF-based multisensor fusion \eqref{update_measurementfusion}.

Firstly, each node operates as an independent fusion center, where the information filtering framework facilitates the distributed reformulation of \eqref{update_measurementfusion} through localized measurement aggregation.\\
\textbf{Assimilation of Covariance:}\\
At time instant $k$, we split $\mR_{k}^{-1} \in \mathbb R^{\tilde{n} \times \tilde{n}}$ into $N$ column blocks $\mR_k^{-1}(*j) \in \mathbb R^{\tilde{n} \times n_{j}}$ (the $j$-th column block of $\mR_{k}^{-1}$) and introduce pseudo inversion regarding the full row rank matrix $\nabla \vh_{k}^{j}$ (satisfying $(\nabla \vh_{k}^{j\dagger})^{\rm T}\cdot(\nabla \vh_{k}^{j})^{\rm T}=\mE_{n_{j}\times n_{j}}$, $\mE_{n_{j}\times n_{j}}$ is the identity matrix \cite{Calculatingthe})
\begin{subequations}
    \begin{align}
        &\mP_{k|k}^{-1}=\mP_{k|k-1}^{-1}+\nabla \vh_{k}^{\rm T}\cdot\mR_{k}^{-1}\cdot\nabla \vh_{k},\\
        &\mP_{k|k}^{-1}=\mP_{k|k-1}^{-1}+\nabla \vh_{k}^{\rm T}\cdot\sum_{j=1}^N{\mR_{k}(\ast j)^{-1}\mR_{k}^{j}\cdot(\nabla \vh_{k}^{j\dagger})^{\rm T}}\notag\\
        &\qquad\quad\left( \nabla \vh_{k}^{j} \right) ^{\rm T}\cdot\left( \mR_{k}^{j} \right)^{-1}\cdot\nabla \vh_{k}^{j}, \label{eq_a_1}
    \end{align}
\end{subequations}
Use the fact $\nabla \vh_{k}^{j}=\nabla \vc_{k}^{j}\cdot\mT_{k}^{j}$ ($rank(\mT_{k}^{j})\geq n_{j}$ since $rank(\nabla \vh_{k}^{j})=n_{j}$), \ref{eq_a_1} could be transformed into 
\begin{subequations}
    \begin{align}
        &\mP_{k|k}^{-1}=\mP_{k|k-1}^{-1}+\nabla \vh_{k}^{\rm T}\cdot\sum_{j=1}^N{\mR_{k}(\ast j)^{-1}\mR_{k}^{j}\cdot(\nabla \vh_{k}^{j\dagger})^{\rm T}}\notag\\
        &\qquad\quad (\mT_{k}^{j})^{\rm T}\cdot(\nabla \vc_{k}^{j}) ^{\rm T}\cdot\left( \mR_{k}^{j} \right)^{-1}\cdot \nabla \vc_{k}^{j}\cdot\mT_{k}^{j},\\
        &\mP_{k|k}^{-1}=\mP_{k|k-1}^{-1}+\nabla \vh_{k}^{\rm T}\cdot\sum_{j=1}^N{\mR_{k}(\ast j)^{-1}\mR_{k}^{j}\cdot(\nabla \vh_{k}^{j\dagger})^{\rm T}}\notag\\
        &\qquad\quad (\mT_{k}^{j})^{\rm T}\hat{\mI}_{k}^{j}\mT_{k}^{j}. \label{eq_a_2}
    \end{align}
\end{subequations}
where local error information matrix has the following expression \cite{chong1990distributed}
\begin{align}
\hat{\mI}_{k}^{j}&=\left( \nabla \vc_{k}^{j} \right) ^{\rm T}\cdot\left( \mR_{k}^{j} \right)^{-1}\cdot \nabla \vc_{k}^{j},\notag\\
&=\left(\mP_{k|k}^{j}\right)^{-1}-\left(\mP_{k|k-1}^{j}\right)^{-1}.
\end{align}
Then the global covariance assimilation expression \eqref{eq_a_2} corresponds to (\ref{global DIF with CCMN_P} - \ref{global fusion weight}). This expression is placed at each node (decentralized) by premultiplying by $(\mT_{k}^{i \dagger})^{\rm T}$ and postmultiplying by $\mT_{k}^{i \dagger}$ to yield the distributed and decentralized nodal covariance assimilation equation \ref{local DIF with CCMN_P}:
\begin{small}
    \begin{subequations}
    \begin{align}
         (\mT_{k}^{i \dagger})^{\rm T}\mP_{k|k}^{-1}\mT_{k}^{i \dagger}&=(\mT_{k}^{i \dagger})^{\rm T}\mP_{k|k-1}^{-1}\mT_{k}^{i \dagger}+ (\mT_{k}^{i \dagger})^{\rm T}\nabla \vh_{k}^{\rm T}\sum_{j=1}^N{\mR_{k}(\ast j)^{-1}} \notag\\
         &\quad\,\,\mR_{k}^{j}\cdot (\nabla \vh_{k}^{j\dagger})^{\rm T}\cdot (\mT_{k}^{j})^{\rm T}\hat{\mI}_{k}^{j}\mT_{k}^{j}\mT_{k}^{i \dagger},\\
        (\mP_{k|k}^{i})^{-1}&=(\mP_{k|k-1}^{i})^{-1}+(\mT_{k}^{i \dagger})^{\rm T}\cdot\nabla \vh_{k}^{\rm T}\cdot\sum_{j=1}^N{\mR_{k}(\ast j)^{-1}}\notag\\
        &\qquad\quad \mR_{k}^{j}\cdot(\nabla \vh_{k}^{j\dagger})^{\rm T}\cdot(\mT_{k}^{i})^{\rm T}(\mT_{k}^{ij})^{\rm T}\hat{\mI}_{k}^{j}\mT_{k}^{ij}. \label{eq_a_3}
    \end{align}
\end{subequations}
\end{small}
It is clear that \ref{eq_a_3} is consistent with (\ref{local DIF with CCMN_P}-\ref{local fusion weight}).\\
\textbf{Assimilation of State:}\\
Derivation of the state estimate assimilation equations is similar to  covariance assimilation equations.
\begin{subequations}
\begin{align}
&(\mP_{k|k})^{-1}\hat{\vx}_{k|k}=\mP_{k|k-1}^{-1}\hat{\vx}_{k|k-1}+\nabla \vh_{k}^{\rm T}\cdot\mR_{k}^{-1}\tilde{\vz}_{k}, \\
&(\mP_{k|k})^{-1}\hat{\vx}_{k|k}=\mP_{k|k-1}^{-1}\hat{\vx}_{k|k-1}+\nabla \vh_{k}^{\rm T}\cdot\sum_{j=1}^N{\mR_{k}(\ast j)^{-1}\mR_{k}^{j}}\notag\\
&\qquad\qquad\qquad\quad(\nabla \vh_{k}^{j\dagger})^{\rm T}\left( \nabla \vh_{k}^{j} \right) ^{\rm T}\cdot\left( \mR_{k}^{j} \right)^{-1}\cdot\tilde{\vz}_{k}^{j},\\
&(\mP_{k|k})^{-1}\hat{\vx}_{k|k}=\mP_{k|k-1}^{-1}\hat{\vx}_{k|k-1}+\nabla \vh_{k}^{\rm T}\cdot\sum_{j=1}^N{\mR_{k}(\ast j)^{-1}\mR_{k}^{j}}\notag\\
&\qquad\qquad\qquad\quad(\nabla \vh_{k}^{j\dagger})^{\rm T}(\mT_{k}^{j})^{\rm T}\cdot(\nabla c_{k}^{j})^{\rm T}\cdot\left( \mR_{k}^{j} \right)^{-1}\cdot\tilde{\vz}_{k}^{j},\\
&(\mP_{k|k})^{-1}\hat{\vx}_{k|k}=\mP_{k|k-1}^{-1}\hat{\vx}_{k|k-1}+\nabla \vh_{k}^{\rm T}\cdot\sum_{j=1}^N{\mR_{k}(\ast j)^{-1}\mR_{k}^{j}}\notag\\
&\qquad\qquad\qquad\quad(\nabla \vh_{k}^{j\dagger})^{\rm T}(\mT_{k}^{j})^{\rm T}\hat{\vi}_{k}^{j}. \label{eq_a_4}
\end{align}
\end{subequations}
where local error information has the following expression \cite{chong1990distributed},
\begin{align}
\hat{\vi}_{k}^{j}&=\left( \nabla \vc_{k}^{j} \right) ^{\rm T}\cdot\left( \mR_{k}^{j} \right)^{-1}\tilde{z}_{k}^{j},\notag\\
&=\left(\mP_{k|k}^{j}\right)^{-1}\vx_{k|k}^{j}-\left(\mP_{k|k-1}^{j}\right)^{-1}\vx_{k|k-1}^{j}.
\end{align}
Then the global state assimilation expression \eqref{eq_a_4} corresponds to (\ref{global DIF with CCMN_x}, \ref{global fusion weight}). Finally, distributed state estimate assimilation equation can be decentralized \eqref{eq_a_4},
\begin{small}
    \begin{subequations}
    \begin{align}
    &(\mP_{k \mid k}^{i})^{-1}\hat{\vx}_{k \mid k}^{i} =(\mP_{k \mid k-1}^{i})^{-1}\hat{\vx}_{k \mid k-1}^{i}+ (\mT_{k}^{i\dagger})^{\rm T} \nabla \vh_{k}^{\rm T}\cdot\sum_{j=1}^N{\mR_{k}(\ast j)^{-1}}\notag\\
    &\qquad\qquad\qquad\quad\mR_{k}^{j}\cdot(\nabla \vh_{k}^{j\dagger})^{\rm T}(\mT_{k}^{i})^{\rm T}(\mT_{k}^{ij})^{\rm T}\hat{\vi}_{k}^{j}. 
    \end{align}
\end{subequations}
\end{small}

Finally, the conclusion is that when measurement noises $\vw_{k}^{j}$ are cross-correlated, $\nabla \vh_{k}^{j}$ is of full row rank ($rank(\nabla \vh_{k}^{j})=n_{j}$) and the rank of $\mT_{k}^{j}$ is no less than $n_{j}$, the decentralized EIF-based multisensor fusion with CCMN given by the following (\ref{appendix global DIF with CCMN}-\ref{appendix global fusion weight}) is optimal in the sense of being equivalent to the EIF-based multisensor centralized fusion (\ref{predict_measurementfusion}-\ref{update_measurementfusion}).  Meanwhile, the local estimates given by (\ref{appendix local DIF with CCMN}-\ref{appendix local fusion weight}) are consistent for each node in the sensor network.
\begin{itemize}
    \item Global space: 
    \begin{subequations}
    \label{appendix global DIF with CCMN}
    \begin{align}
    &\hat{\vx}_{k \mid k} =\mP_{k \mid k}[\mP_{k \mid k-1}^{-1}\hat{\vx}_{k \mid k-1}+\sum_{j=1}^N \mM_{k}^{j}(\mT_{k}^{j})^{\rm T}\hat{\vi}_k^j] ,\label{appendix global DIF with CCMN_x}\\
    &\mP_{k \mid k}^{-1}=\mP_{k \mid k-1}^{-1}+\sum_{j=1}^N \mM_{k}^{j}(\mT_{k}^{j})^{\rm T}\hat{\mI}_k^j\mT_{k}^{j},\label{appendix  global DIF with CCMN_P}
    \end{align}
    \end{subequations}
    where
    \begin{equation}
        \mM_{k}^{j}=\nabla \vh_{k}^{\rm T}\cdot\mR_{k}^{-1}\left( \ast j \right) \mR_{k}^{j}\cdot\left( \nabla \vh_{k}^{j^\dagger} \right) ^{\rm T}. \label{appendix  global fusion weight}
    \end{equation}
    \item Local subspace:
    \begin{subequations}
     \label{appendix local DIF with CCMN}
    \begin{align}
    &\hat{\vx}_{k \mid k}^{i} =\mP_{k \mid k}^{i}[(\mP_{k \mid k-1}^{i})^{-1}\hat{\vx}_{k \mid k-1}^{i}+\sum_{j=1}^N \tilde{\mM}_{k}^{j}(\mT_{k}^{ij})^{\rm T}\hat{\vi}_k^j] ,\label{appendix local DIF with CCMN_x}\\
    &(\mP_{k \mid k}^{i})^{-1}=(\mP_{k \mid k-1}^{i})^{-1}+\sum_{j=1}^N \tilde{\mM}_{k}^{j}(\mT_{k}^{ij})^{\rm T}\hat{\mI}_k^j\mT_{k}^{ij},\label{appendix local DIF with CCMN_P}
    \end{align}
    \end{subequations}
    where fusion weights switch to 
    \begin{equation}
        \tilde{\mM}_{k}^{j}=(\mT_{k}^{i\dagger})^{\rm T}\cdot \nabla \vh_{k}^{\rm T}\cdot\mR_{k}^{-1}\left( \ast j \right) \mR_{k}^{j}\cdot\left( \nabla \vh_{k}^{j^\dagger} \right) ^{\rm T}\cdot(\mT_{k}^{i})^{\rm T}. \label{appendix local fusion weight}
    \end{equation}
\end{itemize}
\twocolumn
\bibliography{reference}

\end{document}